\documentclass[prd,showpacs,twocolumn,superscriptaddress,nofootinbib,floatfix,10pt]{revtex4-2}
\usepackage{setspace}
\usepackage[utf8]{inputenc}
\usepackage{makecell}
\usepackage{float}
\usepackage{amsmath,amssymb,amsfonts,bm}
\usepackage{graphicx}
\usepackage{multirow}
\usepackage{xcolor}
\definecolor{poscolor} {RGB} {252,188,190} 
\definecolor{negcolor} {RGB} {168,168,234} 
\usepackage{slashed}
\usepackage[normalem]{ulem}

\usepackage{colortbl}
\allowdisplaybreaks 
\usepackage[
colorlinks=true,
linkcolor=blue,
breaklinks=true,
citecolor=blue]{hyperref}
\usepackage[figuresright]{rotating}

\colorlet{Miguel}{red!40!orange}

\newcommand{\itp}{\affiliation{CAS Key Laboratory of Theoretical Physics, Institute of Theoretical Physics,\\ Chinese Academy of Sciences, Beijing 100190, China}}

\newcommand{\ucas}{\affiliation{School of Physical Sciences, University of Chinese Academy of Sciences, Beijing 100049, China}}

\newcommand{\ific}{\affiliation{Instituto de F\'isica Corpuscular (centro mixto CSIC-UV),
Institutos de Investigaci\'on de Paterna, \\ Apartado 22085, 46071 Valencia, Spain
}}

\newcommand{\imp}{\affiliation{Institute of Modern Physics, Chinese Academy of Sciences, Lanzhou 730000, China}}

\newcommand{\uestc}{\affiliation{School of Physics, University of Electronic Science and Technology of China, Chengdu 611731, China}}

\newcommand{\peng}{\affiliation{Peng Huanwu Collaborative Center for Research and Education, Beihang University, Beijing 100191, China}}

\begin{document}

\title{\boldmath Understanding the $0^{++}$ and $2^{++}$ charmonium(-like) states near 3.9 GeV}

\author{Teng~Ji}\email{jiteng@itp.ac.cn}
\itp \ucas

\author{Xiang-Kun~Dong}\email{dongxiangkun@itp.ac.cn}
\itp \ucas

\author{Miguel~Albaladejo}\email{Miguel.Albaladejo@ific.uv.es}
\ific

\author{Meng-Lin~Du}\email{du.ml@uestc.edu.cn}
\uestc\ific

\author{Feng-Kun~Guo}\email{fkguo@itp.ac.cn}\itp \ucas \peng 

\author{Juan~Nieves}\email{jmnieves@ific.uv.es}
\ific

\author{Bing-Song~Zou}\email{zoubs@itp.ac.cn}\itp \ucas \imp


\begin{abstract} 
  We propose that the $X(3915)$ observed in the $J/\psi\,\omega$ channel is the same state as the $\chi_{c2}(3930)$, and the  $X(3960)$, observed in the $D_s^+D_s^-$ channel, is an $S$-wave $D_s^+ D_s^-$ hadronic molecule. In addition, the $J^{PC}=0^{++}$ {component in the $B^+\to D^+D^-K^+$} assigned to the $X(3915)$ in the current {\it Review of Particle Physics} has the same origin as the $X(3960)$, which has a mass around 3.94~GeV.
  To check the proposal, the available data in the $D\bar D$ and $D_s^+ D_s^-$ channels from both $ B$ decays and $\gamma\gamma$ fusion reaction are analyzed considering both the $D\bar D$-$D_s\bar D_s$-$D^*\bar D^*$-$D_s^*\bar D_s^*$ coupled channels with $0^{++}$ and a $2^{++}$ state introduced additionally. 
  It is found that all the data in different processes can be simultaneously well reproduced, and the coupled-channel dynamics produce four hidden-charm scalar molecular states with masses around 3.73, 3.94, 3.99 and 4.23~GeV, respectively.
  The results may deepen our understanding of the spectrum of charmonia as well as of the interactions between charmed hadrons.
\end{abstract}

\maketitle

\section{Introduction}
In the last two decades, { many} charmonium(-like) states have been observed in experiments in the charmonium region, which have significantly enriched the hadron spectrum and extended the traditional $c\bar c$ mesons to the so-called exotic states, see Refs.~\cite{Chen:2016qju,Hosaka:2016pey,Richard:2016eis,Lebed:2016hpi,Esposito:2016noz,Guo:2017jvc,Ali:2017jda,Olsen:2017bmm,Belle-II:2018jsg,Cerri:2018ypt,Liu:2019zoy,Brambilla:2019esw,Guo:2019twa,Yang:2020atz,Dong:2021juy,Chen:2022asf,Dong:2021bvy} for recent reviews. The masses of charmonia have been {calculated} in, e.g., Ref.~\cite{Godfrey:1985xj}, among which the low-lying ones are in good agreement with experimental results, but the highly excited states do not match the experimental pattern. 

Focusing on the energy region near 3.9 GeV, there are several experimental candidates of charmonium(-like) states, {namely, the $X(3915)$, the $Z(3930)$, the $\chi_{c0}(3930)$, the $\chi_{c2}(3930)$, and the $X(3960)$.} The $X(3915)$ was observed in the $J/\psi\,\omega$ final state first in the $B$ decays by Belle~\cite{Belle:2004lle} and BaBar~\cite{BaBar:2007vxr}, denoted by $Y(3940)$ there, and later in the two-photon fusion reaction by Belle~\cite{Belle:2009and}, but the quantum number possibilities $J^{PC}=0^{++}$ or $2^{++}$ could not be distinguished. Soon after this observation, it was argued that the $X(3915)$ was the $P$-wave charmonium $\chi_{c0}(2P)$  from the analysis of its decay pattern~\cite{Liu:2009fe}. Later, BaBar confirmed the existence of the $X(3915)$ and suggested $J^{PC}=0^{++}$ quantum numbers for this state~\cite{BaBar:2012nxg}. 
But this assignment is in disagreement with the expected properties for the conventional $P$-wave charmonium $\chi_{c0}(2P)$~\cite{Guo:2012tv,Olsen:2014maa,Olsen:2019lcx}. Moreover, it was also argued in Ref.~\cite{Zhou:2015uva} that the helicity-2 dominance hypothesis, which is reasonable for the coupling of a $2^{++}$ $c\bar c$ state to two photons~\cite{Li:1990sx} and it is supported by experimental measurements~\cite{BESIII:2012uyb}, adopted in the BaBar analysis~\cite{BaBar:2012nxg} was not reliable since the $X(3915)$ may not be a purely $c\bar c$ state.  In Ref.~\cite{Ortega:2017qmg}, the authors found that the structure of both $0^{++}$ and $2^{++}$ states in this energy region are dominantly molecular with a probability of bare $q\bar{q}$ states lower than $45\%$ in the framework of a constituent quark model together with the opening of nearby meson-meson channels. If such an assumption was removed, the data appear more consistent with the assignment of $2^{++}$ to the $X(3915)$~\cite{Zhou:2015uva}. For a related discussion on the {possible $2^{++}$ state in this energy region} as a hadronic molecule, see Ref.~\cite{Baru:2017fgv}.\footnote{It was also found in Ref.~\cite{Baru:2017fgv} that the helicity-2 amplitude dominates over the helicity-0 one within the molecular scenario. The conclusion there was based on relating the coupling of the spin-2 $D^*\bar D^*$ molecular state ($X_2$) to $D^*\bar D^*$ to that of the $\chi_{c1}(3872)$, also known as $X(3872)$, to $D\bar D^*$, and recasting both resonances into a spin multiplet superfield. However, the superfield was constructed for heavy quarkonia and a direct use to relate the couplings in the case of hadronic molecules is questionable. If the {$2^{++}$ state nearby} is indeed the $X_2$ state, its binding energy would be much larger than that of the $\chi_{c1}(3872)$ as a $D\bar D^*$ molecule, and the coupling $g_{X_2 D^*\bar D^*}$ would naturally also be much larger than $g_{\chi_{c1}(3872) D\bar D^*}$.} Besides, both the Belle~\cite{Belle:2005rte} and BaBar~\cite{BaBar:2010jfn} experiments reported a peak structure, denoted as $Z(3930)$ originally, in the $\gamma\gamma\to D \bar D$ reactions near 3.93 GeV. Since the measured helicity angular distributions suggested a spin-2 state, it was then assigned to the radially excited charmonium $\chi_{c2}(2P)$, and is now denoted as $\chi_{c2}(3930)$ in the {\it Review of Particle Physics} (RPP)~\cite{ParticleDataGroup:2022pth}. Recently, the LHCb Collaboration reported  measurements for the $B^+\to K^+D^+ D^-$~\cite{LHCb:2020pxc} and $B^+\to K^+D_s^+ D_s^-$~\cite{LHCb:2022vsv,LHCb:2022dvn} decays. In the $D^+ D^-$ channel, it was found that both $0^{++}$ [$\chi_{c0}(3930)$] and $2^{++}$ [$\chi_{c2}(3930)$] are needed to describe the structure near 3.93 GeV.  {The baseline fit of Ref.~\cite{LHCb:2020pxc} shows that the two states have similar contributions around 3.93~GeV. A comparison of the $\chi^2$ values from different fits is presented in Table VII of that reference. 
One can see there that variations in the region of 3.93~GeV, including either one of the $\chi_{c0}(3930)$ and $\chi_{c2}(3930)$ resonances, produce increases of $\chi^2$ by at most 30\% from the value of $\chi^2=86.1$ obtained in  the preferred fit. 
Given the large larger number of data-points, such bigger $\chi^2$'s would still lead to acceptable fits, and therefore the origin of the 3.93~GeV peak in the LHCb data has not been settled in our opinion.} In the $D_s^+ D_s^-$ channel, an abrupt enhancement appears just above the $D_s^+ D_s^-$ threshold, which is described by a Flatt\'e-like function,  referred as  $X(3960)$ in Ref.~\cite{LHCb:2022vsv}. The mass and width parameters in the Flatt\'e parameterization are $M_0=(3956 \pm 5 \pm 10) ~\mathrm{MeV}$ and $\Gamma_0=(43 \pm 13 \pm 8) ~\mathrm{MeV}$, respectively. 
These values do not correspond to the pole position ($M - i\Gamma/2$) of the resonance.
The abrupt enhancement just above threshold indicates a nearby pole~\cite{Dong:2020hxe}, as in the $Z_c(3900)$ case \cite{Albaladejo:2015lob,Albaladejo:2016jsg,Pilloni:2016obd,Du:2022jjv}, and it was found in Ref.~\cite{Ji:2022uie} that the signal for the $X(3960)$ can be well described by a bound or virtual state of $D_s^+ D_s^-$ below threshold.

It is still under debate what is the nature of these charmonium(-like) states near 3.9~GeV. In Ref.~\cite{Duan:2020tsx}, the $X(3915)$ and the $Z(3930)$ are considered as the conventional $P$-wave charmonium states, $\chi_{c0}(2P)$ and $\chi_{c2}(2P)$, respectively. The results of Ref.~\cite{Ortega:2017qmg} favor the hypothesis that $X(3915)$ and $X(3930)$ resonances arise as different decay mechanisms of the same $J^{PC}=2^{++}$ state. In Ref.~\cite{Chen:2012wy}, the $\gamma\gamma\to D \bar D$ data in the Belle and BaBar measurements are described by introducing both the $\chi_{c0}(2P)$ and $\chi_{c2}(2P)$ states. In addition, there are also studies that predict  non-conventional charmonium-like states near this region. The $X(3915)$ was interpreted as a $c\bar cs\bar s$ tetraquark state~\cite{Lebed:2016yvr,Wang:2016ujn,Chen:2017dpy,Wang:2016gxp} or a $D_s^+ D_s^-$ molecule~\cite{Li:2015iga}. In the lattice quantum chromodynamics (QCD) study of Ref.~\cite{Prelovsek:2020eiw}, there appears a state near the $D_s^+D_s^-$ threshold that originates from the $D\bar D$-$D_s^+D_s^-$ interaction and couples predominantly to the $D_s^+D_s^-$ channel. The authors point out that it may correspond to the $X(3915)$ and $\chi_{c0}(3930)$, which are suggested to be the same state, while in Ref.~\cite{Duan:2021bna}, the $\chi_{c0}(3930)$ is assigned to the $\chi_{c0}(2P)$.  After the observation of the $X(3960)$, it was interpreted as a molecular state in Refs.~\cite{Ji:2022uie,Bayar:2022dqa,Chen:2022dad,Xin:2022bzt,Xie:2022lyw,Mutuk:2022ckn} and identified as the same state as the $\chi_{c0}(3930)$ in Ref.~\cite{Bayar:2022dqa}, the conventional charmonium $\chi_{c0}(2P)$~\cite{Guo:2022ggl}, or a scalar diquark-antidiquark 
tetraquark state~\cite{Agaev:2022pis,Guo:2022crh}.

{In the current (2022) version of RPP~\cite{ParticleDataGroup:2022pth}, the $X(3915)$ and the $\chi_{c0}(3930)$ are assigned to the same $0^{++}$ state $\chi_{c0}(3915)$; the $Z(3930)$ in $\gamma\gamma\to D\bar D$~\cite{Belle:2005rte,BaBar:2010jfn} and the $2^{++}$ component in $B^+\to K^+D^+D^-$~\cite{LHCb:2020pxc} are assigned to the same $2^{++}$ state $\chi_{c2}(3930)$.} In this work, we assume the following assignments, {which are different from those in RPP~\cite{ParticleDataGroup:2022pth}:}
\begin{itemize}
    \item the $X(3915)$ has the same origin as the $\chi_{c2}(3930)$~\cite{Zhou:2015uva};
    \item the $X(3960)$ would be a $D_s^+ D_s^-$ molecular state~\cite{Ji:2022uie};
    \item the peak structures near 3.93 GeV in the $D^+ D^-$ distribution  {from the $B^+\to K^+D^+D^-$ and in the $D\bar D$ distribution from the $\gamma\gamma\to D\bar D$} would contain two contributions, one from the $2^{++}$ $X(3915)$ and the another one from the molecular state of $D_s^+ D_s^-$, {\it  i.e.}, the $X(3960)$.
\end{itemize}
Under the above assumptions, we construct the corresponding amplitudes to simultaneously fit the experimental $D_{(s)}\bar D_{(s)}$ distributions reported by  Belle~\cite{Belle:2005rte} and BaBar~\cite{BaBar:2010jfn}
for the $\gamma\gamma\to D\bar D$ reaction, and  by LHCb~\cite{LHCb:2020pxc,LHCb:2022vsv}
from the analysis of the $B^+\to K^+D^+ D^-$ and $B^+\to K^+D_s^+ D_s^-$ decays.
\\
\section{Formalism}

\subsection{Interactions between $H\bar H$} \label{sec:HH}
We label the $D\bar D$, $D_s^+ D_s^-$, $D^*\bar D^*$ and $D_s^*\bar D_s^*$, with $(I)J^{PC}=(0)0^{++}$,  channels as 1, 2, 3 and 4, respectively. In the near-threshold region, the interactions can be approximated by considering only contact terms, which at leading order are given by constants.  Heavy quark spin symmetry (HQSS) and light-flavor SU(3) symmetry can be employed to express the contact terms, which read~\cite{Ji:2022uie,Hidalgo-Duque:2012rqv}
\begin{equation}
    V_{ij}=4\sqrt{m_{i1}m_{i2}m_{j1}m_{j2}}\,\tilde V_{ij},
\end{equation}
with
\begin{widetext}
    \begin{align}
      \tilde V= \frac12 \left(
\begin{array}{cccc}
  2\mathcal C_{0a} & \sqrt{2}(\mathcal C_{0a}- \mathcal C_{1a}) &  2\sqrt{3} \mathcal  C_{0b} & \sqrt{6}( \mathcal C_{0b}- \mathcal C_{1b}) \\
 \sqrt{2}(\mathcal C_{0a}- \mathcal C_{1a}) & \mathcal C_{0a}+ \mathcal C_{1a} & \sqrt{6} ( \mathcal C_{0b}- \mathcal C_{1b})& \sqrt{3} ( \mathcal C_{0b}+ \mathcal C_{1b})\\
 2\sqrt{3} \mathcal  C_{0b} & \sqrt{6}( \mathcal C_{0b}- \mathcal C_{1b}) & 2( \mathcal C_{0a}-2  \mathcal C_{0b}) & \sqrt{2}({ \mathcal C_{0a}-2  \mathcal C_{0b}- \mathcal C_{1a}+2
    \mathcal C_{1b}}) \\
 \sqrt{6}( \mathcal C_{0b}- \mathcal C_{1b}) & \sqrt{3}( \mathcal C_{0b}+ \mathcal C_{1b}) & \sqrt{2}(\mathcal C_{0a}-2  \mathcal C_{0b}- \mathcal C_{1a}+2
    \mathcal C_{1b}) &  \mathcal C_{0a}-2  \mathcal C_{0b}+ \mathcal C_{1a}-2  \mathcal C_{1b} \\
\end{array}
\right),
\end{align}    
\end{widetext}
where $i,j=1,\ldots,4$ and $m_{i,1}$($m_{i,2}$) is the mass of the first (second) particle in channel $i$. In addition, $\mathcal C_{0a}$, $\mathcal C_{0b}$, $\mathcal C_{1a}$, and $\mathcal C_{1b}$ are the low energy constants (LECs) in the effective Lagrangian and will be rearranged into $\mathcal C_{0a}$, $\mathcal C_{1a}$, 
 $\mathcal C_{0X}=\mathcal C_{0a}+\mathcal C_{0b}$, and $\mathcal C_{1X}=\mathcal C_{1a}+\mathcal C_{1b}$ for later convenience.\footnote{We do not consider open channels like $\eta_c\eta$ below the $D\bar D$ threshold, which can make the LECs complex. Since the experimental data can already be well reproduced with real LECs, as we will show, introducing imaginary parts into the LECs would only bring into the scheme redundant parameters.} In our previous work~\cite{Ji:2022uie}, these LECs are estimated by four experimental inputs: (i) the pole position of the $\chi_{c1}(3872)$ as an $(I)J^{PC}=(0)1^{++}$ $D\bar D^*$ bound state; (ii) the isospin violation decay ratio of the $\chi_{c1}(3872)$, $\mathcal{B}_{\chi_{c1}(3872)\to J/\psi\pi^+\pi^-}/\mathcal{B}_{\chi_{c1}(3872)\to J/\psi\pi^+\pi^-\pi^0}$; (iii)  the pole position of the $Z_c(3900)$ as an $(I)J^{PC}=(1)1^{+-}$ $D\bar D^*$ virtual state; and (iv) the $X(3960)$ as a $J^{PC}=0^{++}$ virtual or bound state of  $D_s^+ D_s^-$. The first two inputs are much more precise than the latter two, which implies that $\mathcal C_{0X}=-0.73_{-0.02}^{+0.01}$~fm$^2$ and $\mathcal C_{1X}=-0.29_{-0.08}^{+0.06}$~fm$^2$ determined from i) and ii) are more reliable than the other two LECs. Therefore, we will fix $\mathcal C_{0X}$ (the most precise one) in what follows and { ensure that the fitted value of $\mathcal C_{1X}$ is consistent with} the previous result obtained in Ref.~\cite{Ji:2022uie}. 

The scattering amplitudes are given by
\begin{align}
    T(W)=[1-VG(W)]^{-1}V, \label{eq:T}
\end{align}
where $G(W)$ is a diagonal matrix with the nonvanishing matrix element $G_{ii}(W)$ given by the meson two-point loop function,
\begin{align}
    G_{ii}(W)=\int \frac{d^4l}{(2\pi)^4}\frac i{(l^2-m_{i1}^2+i\epsilon)[(P-l)^2-m_{i2}^2+i\epsilon]} ,
    \label{eq:G}
\end{align}
with $m_{i1},m_{i2}$ the masses of the intermediate particles in channel $i$, and $P$ their four-momentum [$P^\mu = (W,\vec 0\,)$ in the center-of-mass (c.m.) frame]. 
Using dimensional regularization (DR), it reads
\begin{align}
   & G_{ii}^{\rm DR}(W)=\frac{ 1}{16\pi^2}\bigg\{a_i(\mu)+\log\frac{m_{i1}^2}{\mu^2}+\frac{s-\Delta_i}{2s} \log\frac{m_{i2}^2}{m_{i1}^2} \nonumber\\
&+\frac{k_i}{W} \Big[ 
\log\left(2k_i W+s+\Delta_i\right) + 
\log\left(2k_i W+s-\Delta_i\right) \nonumber\\ 
& -  
\log\left(2k_i W-s+\Delta_i\right) - 
\log\left(2k_i W-s-\Delta_i\right)
\Big]\bigg\},\label{eq:GDR}
\end{align}
where $s=W^2$, $\Delta_i= m_{i1}^2-m_{i2}^2$, $k_i=\lambda^{1/2}(W^2,m_{i1}^2,m_{i2}^2)/(2W)$ is the corresponding three-momentum magnitude with $\lambda(x,y,z)=x^2+y^2+z^2 - 2xy - 2yz - 2xz$  the K\"all\'en triangle function, and $a(\mu)$ is a subtraction constant with  $\mu$, chosen to be 1 GeV, the DR scale. The branch cut of $k_i$, taken from the threshold of the $i$-th channel to infinity along the positive real $W$ axis, splits the whole complex energy ($W$) plane into $2^4=16$ Riemann sheets (RSs), denoted by $r=(\pm,\pm,\pm,\pm)$ and on each RS, Im$(k_i)=r_i|$Im$(k_i)|$. Another way to regularize the loop integral is to insert a Gaussian form factor, namely,
\begin{align}
 G_{ii}^{\Lambda}(W) =&\, 
 \int \frac{l^2 dl}{2\pi^2}  
 \frac{e^{-2l^2/\Lambda^2}/({4m_{i1}m_{i2}}) }{W-l^2/2\mu_{i}-m_{i1}-m_{i2}+i\epsilon},\label{eq:GGF}
\end{align}
with $\mu_{i}=m_{i1}m_{i2}/(m_{i1}+m_{i2})$ the reduced mass of the two particles in channel $i$, where the nonrelativistic approximation has been taken for both intermediate particles. 
{While the expression in Eq.~\eqref{eq:GDR} behaves well in the whole energy region, the subtraction constant $a(\mu)$ is totally unknown. In contrast, the cutoff of the Gaussian form factor in Eq.~\eqref{eq:GGF} has a natural range of $0.5\sim 1.0$~GeV, though it distorts the energy dependence of the loop function away from the threshold region. Therefore, we will use Eq.~\eqref{eq:GDR} for the loop integral but with the subtraction constant determined by matching the two differently regularized loop integrals at threshold. }
We take $\Lambda=1.0$ GeV in the following analysis. The subtraction constant $a_i(\mu)$ in DR is determined by matching the values of the loop function $G_{ii}$ obtained from these two methods at threshold, $W=(m_{i1}+
m_{i2})$. We will use the DR loops with the so-determined subtraction constants for numerical calculations.

\subsection{Distribution formulas for the LHCb data}\label{sec:LHCb}

The LHCb Collaboration has measured the  $B^+\to D^+D^-K^+$~\cite{LHCb:2020pxc} and $B^+\to D_s^+D_s^-K^+$~\cite{LHCb:2022vsv} decays. The $D^+D^-$ and $D_s^+D_s^-$ pairs can couple both to $J^{PC}=0^{++}$ and $2^{++}$ quantum numbers. 
Therefore, both sets of quantum numbers will be considered in our analysis. We aim to analyze the data with a minimal number of resonances, in the spirit of Occam's razor principle. We assume that the $0^{++}$ signatures in both reactions have the same origin, and the $2^{++}$ component is from the known $\chi_{c2}(3930)$.  In principle, the $\chi_{c2}(3930)$ also contributes to the $D_s^+D_s^-$ distribution. However, the LHCb analysis shows that the $0^{++}$ quantum numbers are preferred over $2^{++}$ ones by $12.3\sigma$, and the inclusion of the $\chi_{c2}(3930)$ into their baseline analysis does not lead to a significant improvement~\cite{LHCb:2022vsv}. In addition, for the decay $B^+\to D_s^+D_s^-K^+$, there are no publicly available angular distribution data. Thus, the $\chi_{c2}(3930)$ will be neglected in the present analysis of the $D_s^+D_s^-$ distribution.

The $2^{++}$ component is parameterized as a Breit-Wigner resonance with mass $m_2$ and width  $\Gamma_2$,
\begin{align}
    \mathcal M_{D}(s,z_B)=H_D\frac{m_2 \Gamma_2}{s-m^2_2+i m_2 \Gamma_2}\frac{1-3z_B^2}2, 
    \label{eq:BW2}
\end{align}
where $z_B=\cos\theta_B$, with $\theta_B$ the helicity angle, defined as the angle between the outgoing $K^+$ and $D^+$ mesons in the c.m. frame of the $D^+D^-$. The overall normalization $H_D$ is a free parameter.

The $0^{++}$ component produced in $B^+$ decays is dominated by the process $\bar b u\to c\bar c \bar s u+q\bar q$, with $q=u,d,s$. The productions of the $0^{++}$ $D^0\bar D^0$, $D^+D^-$ and $D_s^+D_s^-$ are in general different~\cite{Savage:1989ub}. 
Hence we introduce three parameters $P_1^+$, $P_1^0$, and $P_2$ for the point-like production sources of $D^+D^-$, $D^0\bar D^0$, and $D_s^+D_s^-$, respectively, and similarly, $P_3^+$, $P_3^0$, and $P_4$ for $D^{*+}D^{*-}$, $D^{*0}\bar D^{*0}$, and $D_s^{*+}D_s^{*-}$, respectively. {Notice that in fact the $B$-meson decay amplitude is complicated and can be energy dependent. Here, however, we focus on only a reduced energy region and the interesting peaks are rather narrow. Therefore, we expect that the impact of the possible energy dependence in the production vertex should be small.} Therefore, the production amplitudes of $D^+D^-$ and $D_s^+D_s^-$, denoted as $ M_{S1}(s)$ and $M_{S2}(s)$, with the final state rescattering, read
 \begin{align}
    \mathcal M_{S1}(s)=&\,P_1^+ + \frac12(P_1^++P_1^0)G_{11}T_{11}+\frac1{\sqrt2} P_2G_{22}T_{21}\notag\\
    &+\frac12(P_3^++P_3^0)G_{33}T_{31}+\frac1{\sqrt2}P_4G_{44}T_{41},\\
    \mathcal M_{S2}(s)=&\,P_2+P_2G_{22}T_{22}+\frac1{\sqrt2}(P_1^++P_1^0)G_{11}T_{12} \notag\\
    &+\frac1{\sqrt2}(P_3^++P_3^0)G_{33}T_{32} +P_4G_{44}T_{42}.
\end{align}
Note that $P_{3}^{+}$ and $P_{3}^{0}$ only appear under the combination $P_{3}^{+} + P_{3}^{0}$ and hence we introduce $P_{3}=P_{3}^{+} + P_{3}^{0}$ as the parameter to be fitted.

The $D^+D^-$ and $D_s^+D_s^-$ invariant mass distributions are then given by
\begin{align}
    \frac{{ d\Gamma}}{d m_{D^+ D^-}}=&\,\frac{1}{(2\pi)^3}\frac{p_1\, q_1}{8m_B^2}\notag\\
    &\times\int_{-1}^1 dz_B\left(|\mathcal M_{S1}|^2+|\mathcal M_D|^2+{c}\right),\label{eq:invmassdist_DDbar_Bdecay}\\
    \frac{{ d\Gamma}}{d m_{D_s^+ D_s^-}}=&\,H_{S2}\frac{1}{(2\pi)^3}\frac{p_2\, q_2}{4m_B^2}|\mathcal M_{S2}|^2,
\end{align}
and the $D^+D^-$ helicity angular distribution reads
\begin{align}
    \frac{{ d\Gamma}}{d z_B}=
   \int_{\sqrt {s_i}}^{\sqrt{s_f}} d\sqrt s\frac{1}{(2\pi)^3}\frac{p_1\, q_1}{8m_B^2}\left(|\mathcal M_{S1}+e^{i\alpha}\mathcal M_{D}|^2+{c}\right),
\end{align}
where
$[s_i,s_f]=[15,16]\,\rm GeV^{2}$ is the energy region corresponding to the helicity angular distribution. $H_{S2}$ is a normalization constant for the $D_s^+D_s^-$ invariant mass distribution, and $p_{1(2)}$ is the magnitude of the three-momentum of the $K^+$ in the rest frame of the $B^+$ meson for the decay $B^+\to K^+D^+D^-(D_s^+D_s^-)$ and $q_{1(2)}$ is the magnitude of the three-momentum of the $D^+(D_s^+)$ in the c.m. frame of the $D^+D^-(D_s^+D_s^-)$ pair. 
{Here for the $B^+ \to D^+D^-K^+$ reaction, we consider the $D\bar D$ rescattering (the coupled-channel $D^{(*)}\bar D^{(*)}$-$D_s^{(*)}\bar D_s^{(*)}$ system with $0^{++}$)  and the $2^{++}$ resonance.
Rescattering and possible resonances such as the $X(2900)$ in the cross channels ({\it i.e.}, $DK$ and $\bar D K$) are not  explicitly taken into account  While their effects can lead to nontrivial structures in $DK$ and $\bar D K$ invariant mass distributions, their projections to the $D\bar D$ distribution in the region of interest is rather flat (see Fig.11 of Ref.~\cite{LHCb:2020pxc}).
The amplitude of such contributions, which are parameterized into the $P_i$'s and $H_D$ parameters, is complex because $DK$ and $\bar DK$, as well as other channels that couple to them, can go on-shell, and the phases of its projections to different partial waves of $D\bar D$ are not the same. 
Consequently, we have introduced a phase factor $e^{i\alpha}$ to account phenomenologically for such contributions to the interference between the $0^{++}$ and $2^{++}$ components in the angular distribution.}
{This interference vanishes when integrating over $z_B$ in Eq.~\eqref{eq:invmassdist_DDbar_Bdecay}.}  In addition, $c$ is a free parameter introduced as a background without interference, but finally it turns out that from the best fit it is compatible with zero within  errors and therefore, we will drop this parameter in the following analysis.

{The formulas used to fit the experimental data read
\begin{align}
    \Delta N_1&=A_1 \frac{\Delta m_{D^+D^-}^2}{2m_{D^+D^-}} \frac{{ d\Gamma}}{d m_{D^+ D^-}},\\
    \Delta N_2&=A_2\Delta {m_{D_s^+D_s^-}} \frac{{ d\Gamma}}{d m_{D_s^+ D_s^-}},\\
    \Delta N_3&=A_3\Delta z_{B} \frac{{ d\Gamma}}{dz_B},
\end{align}
with $\Delta N_i$ the events in each energy bin and $A_i$ the corresponding normalization constants with unit of MeV$^{-1}$. $\Delta m_{D^+D^-}^2=3.7\times10^4$ MeV$^{2}$, $\Delta m_{D_s^+D_s^-}=20$ MeV and $\Delta z_B=0.067$ are the corresponding bin widths in experiments. Note that $A_1=A_3$ can be absorbed by productions and $A_2$ by $H_{S2}$.}
\subsection{Distribution formulas for the Belle and BaBar data}

The differential cross section of the $\gamma\gamma\to D\bar D$ can be represented by two independent helicity amplitudes $\mathcal M_{+\pm}$ as~\cite{Zhou:2015uva}
\begin{align}
    \frac{d \sigma}{d \Omega}=\frac{\sqrt{s-4m_D^2}}{64 \pi^2  s^{3/2}}\left(\left|\mathcal{M}_{++}\right|^2+\left|\mathcal{M}_{+-}\right|^2+|\mathcal M_{\rm bg}|^2\right),
\end{align}
where
\begin{align}
    &\mathcal{M}_{++}(s, z_\gamma)=16 \pi \sum_{J \geq 0}(2 J+1) F_{J, 0}(s) d_{0,0}^J(z_\gamma),\\
    &\mathcal{M}_{+-}(s, z_\gamma)=16 \pi \sum_{J \geq 2}(2 J+1) F_{J, 2}(s) d_{2,0}^J(z_\gamma),
\end{align}
where $F_{J,0}$ and $F_{J,2}$ are the partial-wave amplitudes with the angular momentum $J$ for helicities 0 and 2, respectively, $z_\gamma=\cos\theta_{\gamma}$, with $\theta_{\gamma}$ the angle between the outgoing $D$ and the incoming $\gamma$ in the c.m. frame, and $d^J_{m,m'}$ are the Wigner $d$-functions. In addition, {we have introduced a $0^{++}$ Breit-Wigner resonance as the noninterfering background,}
\begin{align}
    \mathcal M_{\rm bg}&= H_{\rm bg} \frac{m_0\Gamma_0(s)}{s-m_0^2+im_0\Gamma_0(s)}, \notag\\ \Gamma_0(s)&=\Gamma_0\frac{p(s)}{p(m_0^2)}\frac{m_0}{\sqrt s}, \label{eq:BW0}
\end{align}
where $p(s)$ is the magnitude of the three-momentum of $D^+$ in the c.m. frame of the $D^+D^-$, and $m_0,\Gamma_0$ are two free parameters.

We consider only the lowest partial waves, $F_{0,0},F_{2,0}$ and $F_{2,2}$, which are expressed as
\begin{align}
    F_{00}=&\,P_{\gamma1}+P_{\gamma1}G_{11}T_{11}+P_{\gamma2}G_{22}T_{21}\notag\\
    &+P_{\gamma3}G_{33}T_{31}+P_{\gamma4}G_{44}T_{41},\\
    F_{20}=&\,H_{\gamma0}m_2\Gamma_2/(s-m_2^2+im_2\Gamma_2),\\
    F_{22}=&\,H_{\gamma2}m_2\Gamma_2/(s-m_2^2+im_2\Gamma_2),
\end{align}
with $P_{\gamma i}$ the production of channel $i$ in the $\gamma\gamma$ annihilation reaction. We have tried many different fits, and it turns out that the four production parameters $P_{\gamma i}$ have almost the same absolute values and hence they are represented by the same parameter $P_{\gamma}$ in the final result to be presented below, with $P_{\gamma 1,2}=-P_{\gamma 3,4}=P_{\gamma}$. In addition, $H_{\gamma 0}$ and $H_{\gamma2}$ are two free parameters and represent the contribution of the $2^{++}$ resonance for different helicities, while $m_2$ and $\Gamma_2$ are the same parameters as those in the formulas for the LHCb data [see Eq.~\eqref{eq:BW2}]. The energy distribution is given by
\begin{align}
    \sigma(E)=2\pi{\frac{f_{\rm BaBar}}{N_{m2}}}\frac12\int_{-1}^1dz_\gamma \frac{d \sigma}{d \Omega},
\end{align}
and the angular distribution reads
\begin{align}
    \frac{d \sigma}{d z_\gamma}=2\pi{\frac{1}{N_{\theta2}}}\frac{1}{E_f-E_i}\int_{E_i}^{E_f}dE \frac{d \sigma}{d \Omega},
\end{align}
where $N_{m2}=4$ and $N_{\theta2}=10$ are the numbers of bins of the energy and angular distributions in the experimental data, and $[E_i,E_f]=[3.91,3.95]$~GeV is the energy region corresponding to the angular distribution. Finally, $f_{\rm BaBar}$ is fixed to 1 when fitting to  Belle data while it is treated as a free parameter when fitting to  BaBar data.

\section{Fit results and discussions}

Given the above distribution formulas, the different datasets can be fitted to by minimising the $\chi^2$ function using the MINUIT algorithm~\cite{James:1975dr,iminuit,iminuit.jl}. In total, we have 23 
parameters, $\Lambda$, $\mathcal C_{0X}$, $\mathcal C_{1X}$, $\mathcal C_{0a}$, $\mathcal C_{1a}$, $P_1^+$, $P_1^0$, $P_2$, $P_3$, $P_4$, $P_{\gamma}$, $H_{\gamma0}$, $H_{\gamma2}$, $H_{D}$, $H_{S2}$, $H_{\rm bg}$, $m_2$, $\Gamma_2$,  $m_0$, $\Gamma_0$, $\alpha$, $c$ and $f_{\rm BaBar}$.
The LEC $\mathcal C_{0X}=-0.73$~fm$^2$ is fixed as discussed above in Section~\ref{sec:HH}. The cutoff parameter $\Lambda$ is fixed to 1~GeV, corresponding to the scale at which the quoted $\mathcal C_{0X}$ value is obtained~\cite{Ji:2022uie}. A variation of the cutoff $\Lambda$ can be almost completely absorbed by the LECs.
As discussed at the end of Section~\ref{sec:LHCb}, $c$ is set to 0. After fixing these parameters, there remain 20 free parameters.

\begin{figure*}
    \centering
    \includegraphics[width=\linewidth]{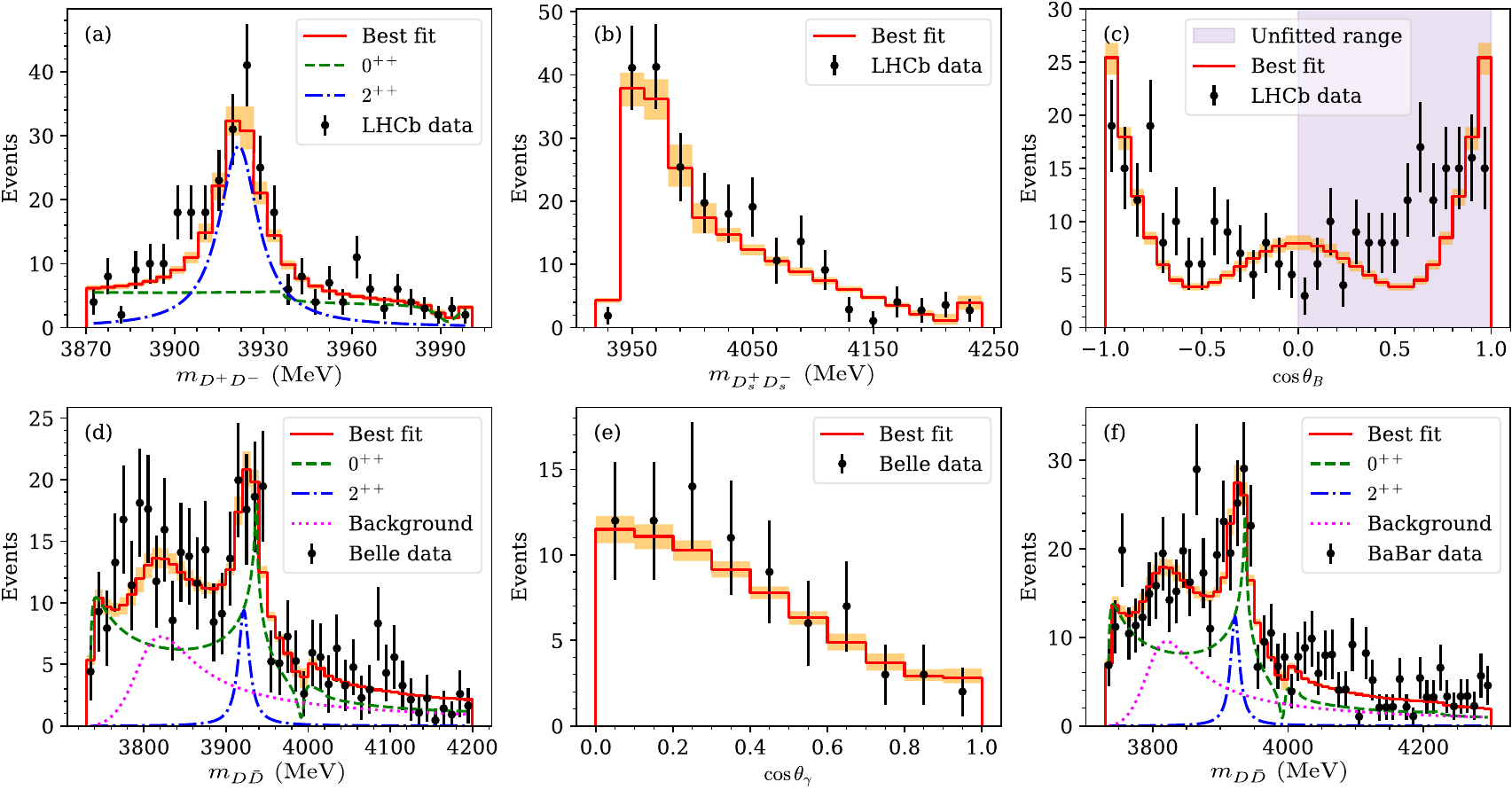}
    \caption{ Best fit to the LHCb data for the $B^+\to K^+D^+D^-$~\cite{LHCb:2020pxc}, $B^+\to K^+D_s^+D_s^-$~\cite{LHCb:2022vsv} decays and the Belle~\cite{Belle:2005rte} and BaBar~\cite{BaBar:2010jfn} data for the $\gamma\gamma\to D\bar D$ reaction. The orange bands stand for the statistical uncertainty of the fitted line shapes, which are inherited from the errors of the different data sets. In each panel, we split the different contributions considered in the present scheme. Note that we have averaged the distributions over each energy bin to compare with the experimental data. The line shapes before averaging are shown in Appendix~\ref{app:para}. }
    \label{fig:fit-lineshape}
\end{figure*}

{There can be multiple solutions to a fitting problem of many parameters. To overcome this issue, we have sampled more than 1000 sets of initial values to fit the data and obtained 75 sets of possible parameters with $\chi^2$/d.o.f.$<1.5$. Most of these parameters lead to unreasonable lineshapes (like sharp kinks). The lineshapes from the remaining parameter sets show small differences. Then we selected the one with the smallest $\chi^2$/d.o.f. as the best fit.}
All the data can be well described under the picture we are considering. 
The best fitted line shapes are shown in Fig.~\ref{fig:fit-lineshape} with 
\begin{equation}
    \chi^2/\rm d.o.f.=176.5/(173-20)=1.15. \label{eq:chisq}
\end{equation}
Note that the LHCb analysis in Ref.~\cite{LHCb:2020pxc} indicates that the $\cos\theta_B$ angular distribution of the $D^+  D^-$  with $\cos\theta_B>0$, shaded region in Fig.~\ref{fig:fit-lineshape}(c), receives a significant contribution from possible $X(2900)$ states in the $DK$ channel. Thus, we only fit to the $\cos\theta_B$ data in the region of $\cos\theta_B<0$, which are  almost free of such contributions, as shown by the red dotted line in Fig.~11(b) of Ref.~\cite{LHCb:2020pxc}. In addition, we use the fitted parameters to calculate the line shapes of the Belle data with $\cos\theta_\gamma>0$ and $\cos\theta_\gamma<0$. They are shown in Fig.~\ref{fig:checkbelle}. We see that the present scheme provides a very good description of these distributions as well.

\begin{figure}
    \centering
    \includegraphics[width=\linewidth]{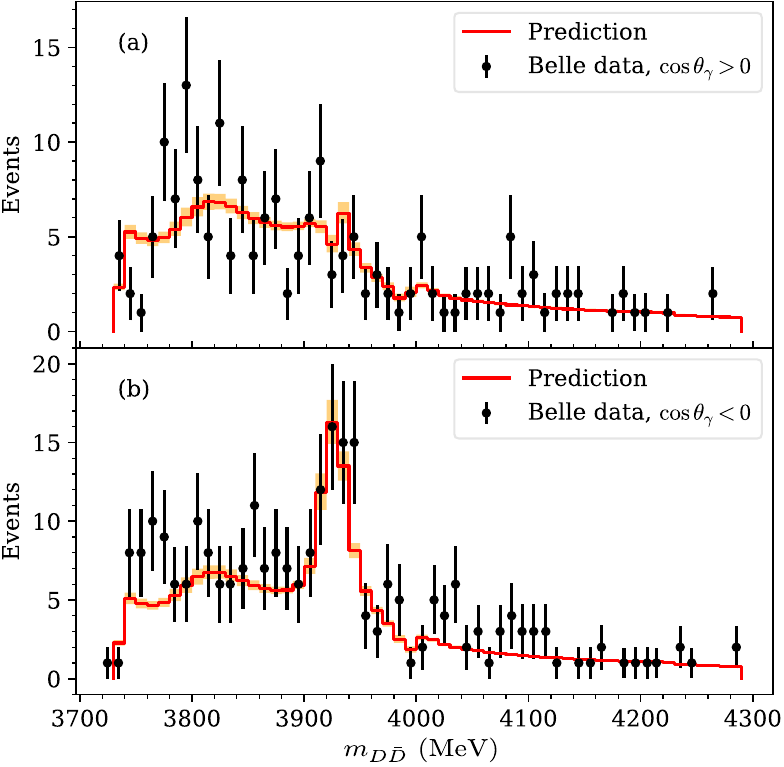}
    \caption{Predicted line shapes for the for the $\gamma\gamma\to D\bar D$ reaction with $\cos\theta_\gamma>0$ (top) and $\cos\theta_\gamma<0$ (bottom). We compare our results with the Belle data~\cite{Belle:2005rte}.  }
    \label{fig:checkbelle}
\end{figure}

The best fit parameters are listed in Appendix~\ref{app:para}. The LECs from the best fit are:\footnote{{{ The error of the fitted parameters, as well as their correlations, are obtained from MINUIT \cite{James:1975dr,iminuit,iminuit.jl}, as mentioned before. The uncertainty from the output quantities computed from these parameters are obtained through Monte Carlo simulations, by sampling the fitted parameters according to multi-variate Gaussian distributions, taking into account their errors and correlations.}}}
\begin{align}
    \mathcal C_{1X}  &=(-0.33\pm0.02)\ \rm fm^2, \notag\\
     \mathcal C_{0a}  &=(-1.36\pm0.06)\ \rm fm^2,\\
     \mathcal C_{1a}  &=(-0.33\pm0.02)\ \rm fm^2. \notag
\end{align}
We see that $\mathcal C_{1X}$, although it is let free in the fit, takes within errors almost the same value as the previously determined one, $\left(-0.29_{-0.08}^{+0.06}\right) \mathrm{fm}^2$, from the $\chi_{c1}(3872)\to \pi^+\pi^-(\pi^0)$ decays~\cite{Ji:2022uie}. 
On the other hand, $ \mathcal C_{1a}$ also agrees quite well with that in Ref.~\cite{Ji:2022uie}, $\left(-0.31_{-0.05}^{+0.03}\right) \mathrm{fm}^2$, while $\mathcal C_{0a}$ here is close to the value, $\left(-1.50_{-0.15}^{+0.17}\right) \mathrm{fm}^2$,  found for scenario II of Ref.~\cite{Ji:2022uie}, where $D_s^+D_s^-$ forms a bound state in the single-channel treatment. In the present framework, there is a virtual-state-like pole below the $D_s^+D_s^-$ threshold (see below) and this difference results from the consideration of coupled-channel dynamics.

The resonance parameters for the $2^{++}$ state are also free in the fit and the obtained mass and width are 
\begin{align}
    m_2=(3922\pm2)\,{\rm MeV},\quad \Gamma_2=(16\pm3)\,{\rm MeV}.
\end{align}
They are consistent with the values for the $X(3915)$ state in the $J/\psi\omega$ mode from the Belle~\cite{Belle:2004lle,Belle:2009and} and BaBar~\cite{BaBar:2010wfc,BaBar:2012nxg} measurements, which were averaged to be $(3918.4 \pm 1.9)$~MeV and $(20 \pm 5)$~MeV, in the 2020 version of RPP~\cite{ParticleDataGroup:2020ssz}. 

The resonance parameters for the $0^{++}$ background in the $\gamma\gamma\to D\bar D$ annihilation reaction are found to be
\begin{align}
    m_0=(3815\pm10)\,{\rm MeV},\quad \Gamma_0=(90\pm13)\,{\rm MeV}.
\end{align}
This background, if interpreted as a resonance, may correspond to the $\chi_{c0}(2P)$ state as argued in Ref.~\cite{Guo:2012tv}. Actually, its mass and width were determined to be $m = (3837.6 \pm 11.5)~\mathrm{MeV}$ and $\Gamma=(221 \pm 19)~\mathrm{MeV}$ in Ref.~\cite{Guo:2012tv}  from the $\gamma\gamma\to D\bar D$ data~\cite{Belle:2005rte,BaBar:2010jfn}, and $m = \left(3862^{+50}_{-35}\right)~\mathrm{MeV}$ and $\Gamma=\left(201^{+180}_{-110}\right)~\mathrm{MeV}$ by the Belle Collaboration in Ref.~\cite{Belle:2017egg} from the $e^+e^-\to J/\psi D\bar D$. 
However, we cannot claim in the present analysis that the background must come {exclusively} from such a resonance, because there could be other sources. In what follows, we consider two other possible scenarios for the background. In the first one, we  introduce the contribution from the off-shell $\chi_{c0}(1P)$, whose mass and width are taken as $3414$~MeV and $10.8$~MeV~\cite{ParticleDataGroup:2022pth}. Although this state is a few hundred MeV below the energy region we are focusing on, the tail of $\chi_{c0}(1P)$ together with an $s$-channel off-shell form factor, following the framework of Ref.~\cite{Liang:2004sd}, provides a description of the data with $\chi^2/\text{d.o.f.}=1.25$, which is only slightly larger than the one reported above with Eq.~\eqref{eq:BW0} as the background.
In the second case, the background term is dropped and the best fit leads to a $\chi^2/\text{d.o.f.}=1.39$. 
The comparison is shown in Fig.~\ref{fig:compare}. 
More precise data are needed to distinguish the different scenarios and to establish the existence or not of the broad $0^{++}$ meson with a mass around 3.8~GeV.

\begin{figure}
    \centering
    \includegraphics[width=\linewidth]{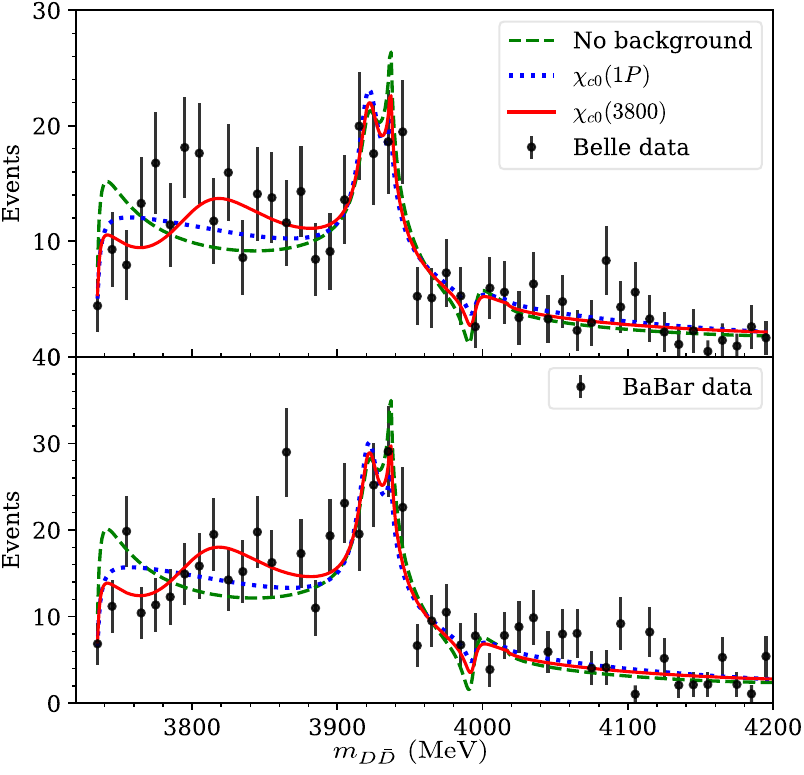}
    \caption{Line shapes of the $\gamma \gamma \to D \bar D$ from the best fits with different parameterizations of the background. The red curves denoted as $\chi_{c0}(3800)$ consider a $0^{++}$ Breit-Wigner resonance (Eq.~\eqref{eq:BW0}) for the background, and they correspond to those from the main result reported in this work ($\chi^2/\rm d.o.f.=1.15$). The blue dotted and green dashed curves show the results obtained using the $\chi_{c0}(1P)$ supplemented with an off-shell form factor ($\chi^2/\rm d.o.f.=1.25$) and those obtained without any background ($\chi^2/\rm d.o.f.=1.39$), respectively; see text for details.}
    \label{fig:compare}
\end{figure}

The found value for the ratio $H_{\gamma2}/H_{\gamma0}=0.8\pm0.7$ supports that the helicity-0 amplitude provides a significant contribution to the
fusion $\gamma\gamma\to D \bar D$ reaction. It is generally consistent with the arguments in Ref.~\cite{Zhou:2015uva} that the pure helicity-2 assumption in the BaBar analysis is not justified.
 
With the fitted LECs, the pole positions of the $H\bar H$ coupled-channel $J^{PC}=0^{++}$ isoscalar $T$-matrix (Eq.~\eqref{eq:T})  can be obtained and the results are collected in Table~\ref{tab:polepos}.\footnote{There exists a shadow pole located at $3987.0- i3.0$ MeV on the RS $r=(-,+,+,+)$. This pole comes from the coupled-channel dynamics~\cite{Eden:1964zz} and it is a ``shadow" of the pole on the RS $r=(-,-,+,+)$. It is further away from the physical region than this latter pole and it has a smaller impact on the physical line shape. Hence in Table~\ref{tab:polepos}, we report the pole on the RS $r=(-,-,+,+)$.} The couplings between these poles and different channels are characterised by the residues of the $T$-matrix, and determined by
\begin{align}
    g^2_{I,i}=\lim_{E\to E_I}(E^2-E_I^2)T_{ii}(E),
\end{align}
where $i$ and $I$ label the channels and the poles, respectively, $E_I$ is the pole position listed in Table~\ref{tab:polepos}, and the $T$-matrix elements should be computed on the RS where the pole is located. The results lead to the following conclusions.

\begin{table*}[!ht]
    \centering
    \renewcommand\arraystretch{1.6}
    \caption{Pole positions of the $D\bar D$-$D_s\bar D_s$-$D^*\bar D^*$-$D_s^*\bar D_s^*$ coupled-channel $J^{PC}=0^{++}$ isoscalar $T$-matrix and their effective couplings to the relevant channels. }\label{tab:polepos}
    \begin{ruledtabular}
    \begin{tabular}{l|c|c|c|c}
        Pole [MeV] & $ 3727.8^{+0.2}_{-0.3}+i0$ & $3936.5^{+0.4}_{-0.9}+i( 16.1^{+4.2}_{-2.2})$ & $3993.1^{+0.4}_{-0.7}+i( -4.5^{+0.2}_{-0.2})$ & $ 4228.1^{+0.2}_{-0.1}+i( 10.7^{+3.4}_{-2.7})$\\
        \hline
        RS & $(+,+,+,+)$ & $(+,-,+,+)$ & $(-,-,+,+)$ & $(+,+,+,-)$ \\ \hline\hline
        Channel&\multicolumn{4}{c}{Coupling $g_{I,i}$ [GeV]}
        \\\hline
        $D\bar D$ & $ 9.36_{-0.07}^{+0.09}+i0$ &$  4.58_{-0.28}^{+0.43}+i( -2.57_{-0.33}^{+0.21})$ &$  0.48_{-0.03}^{+0.03}+i( -1.65_{-0.04}^{+0.02})$ &$  0.63_{-0.05}^{+0.06}+i( 0.01_{-0.02}^{+0.07})$ \\ \hline
        $D_s^+ D_s^-$ & $ 3.72_{-0.07}^{+0.12}+i0$ &$  3.80_{-0.05}^{+0.04}+i( 10.8_{-0.37}^{+0.53})$ &$  0.27_{-0.04}^{+0.06}+i( -2.07_{-0.05}^{+0.09})$ &$  0.55_{-0.03}^{+0.05}+i( 0.01_{-0.15}^{+0.07})$\\ \hline
        $D^*\bar D^*$ & $ 2.20_{-0.04}^{+0.08}+i0$ &$  2.08_{-0.21}^{+0.51}+i( -4.67_{-0.20}^{+0.27})$ &$  13.85_{-0.07}^{+0.10}+i( 0.79_{-0.03}^{+0.02})$ &$  4.22_{-0.44}^{+0.49}+i( -2.44_{-0.32}^{+0.24})$\\ \hline
       $D_s^{*+} D_s^{*-}$&$ 1.37_{-0.04}^{+0.04}+i0$ &$  1.02_{-0.18}^{+0.26}+i( -2.50_{-0.11}^{+0.06})$ &$  6.01_{-0.13}^{+0.16}+i( 0.51_{-0.01}^{+0.02})$ &$  4.83_{-0.03}^{+0.08}+i( 9.91_{-0.66}^{+0.66})$
    \end{tabular}
    \end{ruledtabular}
\end{table*}

There exists a pole about 7 MeV below the lowest threshold on the physical RS $r=(+,+,+,+)$ and it couples most strongly to the $D\bar D$ channel. This corresponds to a isoscalar $0^{++}$ $D\bar D$ bound state, which has been previously predicted by various phenomenology models~\cite{Wong:2003xk,Zhang:2006ix,Gamermann:2006nm,Liu:2009qhy,Nieves:2012tt,Hidalgo-Duque:2012rqv,Hidalgo-Duque:2012rqv,Dong:2021juy}, and more recently by the lattice QCD calculation of Ref.~\cite{Prelovsek:2020eiw}. The existence of the isoscalar $0^{++}$ $D\bar D$ state has also received
support in Refs.~\cite{Gamermann:2007mu,Wang:2020elp,Deineka:2021aeu} from the analysis of the Belle and BaBar data~\cite{Belle:2007woe,Belle:2005rte,BaBar:2010jfn}.
{Since the $D\bar D$ is the lowest channel that is considered, the pole has a vanishing imaginary part. The existence of lower channels, such as the $\eta_c\eta$, $J/\psi\,3\pi$ and $\chi_{c0}\,2\pi$, should give the pole a finite width.}

The second pole located on the RS $r=(+,-,+,+)$ is close to the $D^+_sD^-_s$ threshold, 3937~MeV, and couples to this channel most strongly. 
This pole is virtual-state-like since it is not connected to the physical axis directly, as can be seen in Fig.~\ref{fig:my_label}. Being shaded by the $D_s^+ D_s^-$ threshold, this pole shows up as a threshold cusp, which produces a dip in the $B^+\to K^+D^+D^-$ decay and a peak in the $\gamma\gamma$ fusion reaction, exactly at the threshold of $D^+_sD^-_s$. It is natural that the pole will give rise to reaction-dependent line shapes due to different production rates of the coupled  channels~\cite{Dong:2020hxe}.
The pole should be located on the real axis below the $D^+_sD^-_s$ threshold on the physical or unphysical RS (it would be on the physical RS if the LECs obtained here were used in a single-channel treatment; see Ref.~\cite{Ji:2022uie}), but here it moves to the complex plane because of the coupling to the other channels. In the lattice study carried out in Ref.~\cite{Prelovsek:2020eiw}, the $D^+_sD^-_s$ forms a bound state in a single channel analysis and it becomes a resonance if the coupling to $D\bar D$ is turned on.  The phenomenological study of Ref.~\cite{Meng:2020cbk} also predicted a bound $D^+_sD^-_s$ state, while the light vector meson exchange approach of Ref.~\cite{Dong:2021juy} yields a virtual state of $D^+_sD^-_s$.

The third pole is located on the RS $r=(-,-,+,+)$, about {25}~MeV below the threshold of $D^*\bar D^*$ and couples most strongly to this channel. Therefore, it corresponds to a  $0^{++}$ bound state of the isoscalar $D^*\bar D^*$ pair, which moves to the complex plane due to the couplings to lower energy channels. Such a pole is close to the physical axis and hence it produces clear imprints in the line shapes of $D\bar D$ invariant mass distributions, reflected by the dips in the $0^{++}$ contributions in Fig.~\ref{fig:fit-lineshape}(d) and (f) and in Fig.~\ref{fig:compare}. However, the energy intervals of the binned experimental data are comparable with or even larger than the width of this pole; thus, such dips are smeared when integrating over the energy intervals. It has been argued~\cite {Hidalgo-Duque:2013pva,Baru:2016iwj} the existence of an isoscalar-scalar state $0^{++}$ $D^*\bar D^*$ as one of the HQSS partners of the $\chi_{c1}(3872)$.
The light vector meson exchange also supports the existence of a $D^*\bar D^*$ bound state with $I(J^{PC})=0(0^{++})$~\cite{Dong:2021juy}.
There have also been predictions of a  $D^*\bar D^*$ bound state with  quantum numbers $I(J^{PC})=0(2^{++})$ with a mass around 4~GeV~\cite{Nieves:2012tt,Guo:2013sya,Hidalgo-Duque:2013pva,Baru:2016iwj,Dong:2021juy}. Because in the distributions there are no apparent nontrivial signatures in this energy region, the  contribution of this latter resonance has not been included in the present analysis.

The fourth pole couples most strongly to the $D_s^*\bar D^*_s$ and it is close to its threshold. Similarly to the second pole, it behaves like a virtual state since it is located on a shaded RS [$r=(+,+,+,-)]$. There have been several calculations identifying the resonance $X(4140)$, first observed by the CDF Collaboration~\cite{CDF:2009jgo}, to  a bound $0^{++}$ or $2^{++}$ $D_s^*\bar D^*_s$ state, with a large binding energy about 80~MeV~\cite{Liu:2009ei,Branz:2009yt,Albuquerque:2009ak,Ding:2009vd,Zhang:2009st,Karliner:2016ith}. However, the recent LHCb measurement~\cite{LHCb:2021uow} favours the quantum numbers of the $X(4140)$ to be $1^{++}$ using a Breit-Wigner representation. In Ref.~\cite{Dong:2020hxe}, the $S$-wave $D_s^*\bar D^*_s$ is more likely to form virtual states with quantum numbers $0^{++}$, $2^{++}$ and $1^{+-}$ instead of bound states.
Such a scenario is consistent with the LHCb data~\cite{LHCb:2021uow} of the $J/\psi\phi$ invariant mass distribution in the $B^+\to J/\psi\phi  K^+$ decay, where a clear dip sits exactly at the $D_s^*\bar D^*_s$ threshold. This could be the signal of a nearby virtual-state-like pole~\cite{Dong:2020hxe}.

We now turn to further consequences that can be drawn from our analysis of the probably exotic state $X(3960)$. The present coupled-channel approach allows us to predict the line shapes produced by the second pole, which predominantly couples  to  $D_s^+D_s^-$, {\it i.e.}, the $X(3960)$, in the  $D \bar D$ invariant mass distributions from $e^+e^-\to \phi/\omega D\bar D$ reactions. 
The off-diagonal $D_s^+D_s^- \leftrightarrow  D^+D^-$ transition guaranties that the $X(3960)$ should show up in the  $D \bar D$ invariant mass distributions. {The role of the off-diagonal term in connection with the $X(3960)$ is examined at length in Ref.~\cite{Bayar:2022dqa}}. Taking into account that the quark contents of the $\phi$ and $\omega$ mesons are $\bar s s$ and $(\bar u u + \bar d d )/\sqrt{2}$, respectively, we expect that the $\phi D_s^+ D_s^-$ and $\omega D\bar D$ productions, from the $\bar c\gamma_\mu c$ vector current, should be larger than those of the $\phi D\bar D$ and the $\omega D_s^+ D_s^-$, since the latter two are suppressed by the  Okubo-Zweig-Iizuka rule while the former two are not. 
Therefore, the main production mechanism for the $\phi D\bar D$ should be through the $e^+e^-\to \phi D_s^+D_s^-\to \phi D\bar D$ chain reaction. 
The $S$-wave contribution to the $D \bar D$ invariant mass distributions in the $e^+e^-\to \phi D\bar D,\ \omega D\bar D$ processes can be estimated by $|P_2G_{22}T_{21}|^2$ and $|P_1+P_1G_{11}T_{11}|^2$ multiplied by the corresponding phase space, respectively, where $T_{ij}$ are the $T$-matrix elements as given by Eq.~\eqref{eq:T}, $G_{ii}$ is given by Eq.~\eqref{eq:G}, and $P_{1}$ and $P_2$ are two parameters for the direct productions of the $e^+e^-\to \omega D\bar D$ and $e^+e^-\to \phi D_s^+D_s^-$, respectively.
The predicted line shapes are shown in Fig.~\ref{fig:eepredict}. We can see that the  $D_s^+ D_s^-$ molecular state shows up as a peak exactly at the $D_s^+ D_s^-$ threshold for the $e^+e^-\to \phi D\bar D$ decay, while it produces a dip just below the $D_s^+ D_s^-$ threshold for the $e^+e^-\to \omega D\bar D$ reaction. 
It is similar to the line shape of the $f_0(980)$ in the $J/\psi \to \phi \pi \pi$ and $J/\psi \to \omega \pi \pi$ processes~\cite{BES:2004mws,BES:2004twe} as discussed in Ref.~\cite{Dong:2020hxe}.

\begin{figure}
    \centering
    \includegraphics[width=\linewidth]{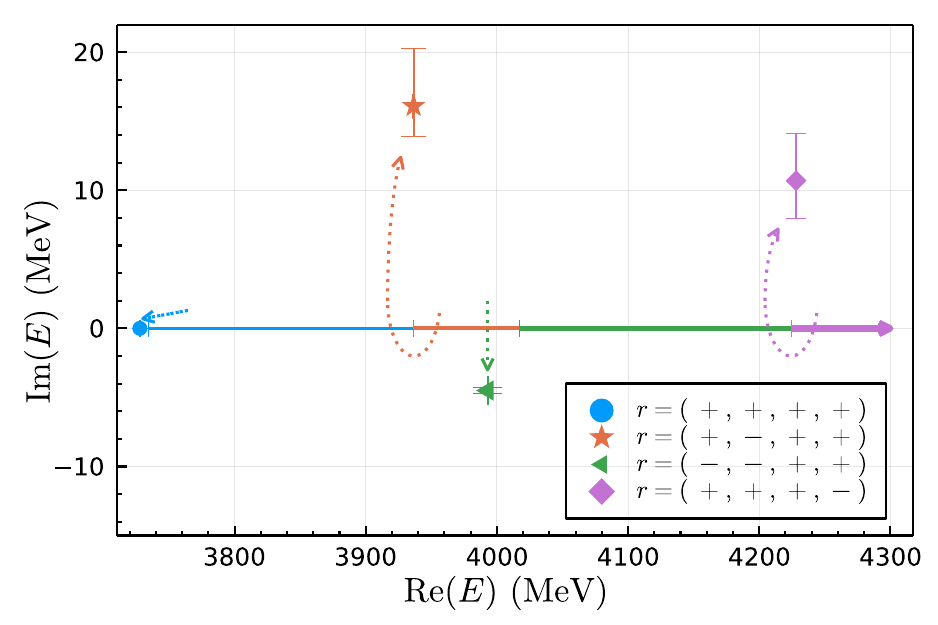}
    \caption{Illustration of the pole positions (points with errors) computed using the best fit parameters and the paths that reach them from the physical region (dotted arrow lines). The horizontal lines with different colors represent the right-hand cuts, which start from the thresholds of the four channels, $D\bar D$, $D_s\bar D_s$, $D^*\bar D^*$ and $D_s^*\bar D^*_s$ from left to right, respectively, to positive infinity. }
    \label{fig:my_label}
\end{figure}
\begin{figure}
    \centering
    \includegraphics[width=\linewidth]{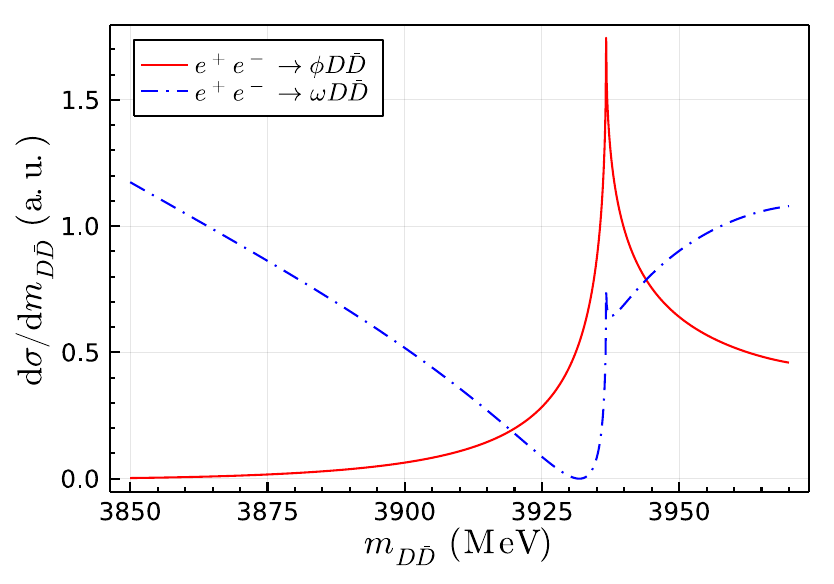}
    \caption{Predicted $D\bar D$ invariant mass distributions, from the $S$-wave contribution near the $D_s^+D_s^-$ threshold due to the existence of the $D_s^+D_s^-$ molecular state, for the $e^+e^-\to \phi/\omega D\bar D$ reactions at $E_{e^+e^-} = 5.4$~GeV.}
    \label{fig:eepredict}
\end{figure}

\vspace{15pt}

\section{Summary}

{The current treatment of the $D\bar D$-$D_s\bar D_s$-$D^*\bar D^*$-$D_s^*\bar D^*_s$ coupled-channel interactions corresponds to a pionless theory at the leading order of the nonrelativistic expansion. In principle, the one-pion exchange (OPE) as a longer-range contribution can also be introduced.
Yet, for the coupled-channel problem at hand with threshold differences of order of 300~MeV, introducing OPE would require the introduction of new counterterms for $S$-$D$-wave mixing to absorb the cutoff dependence.
Since the current data can already be well described, we refrain from doing so.
Nevertheless, the mass spectrum of the coupled channel system is expected to change only marginally.
Such pattern has been seen in previous coupled-channel studies, see, e.g., the $P_c$ mass spectrum without OPE in \cite{Sakai:2019qph} and that with the OPE in \cite{Du:2019pij} or the $D^{(*)}\bar D^{(*)}$ systems without and with OPE in \cite{Nieves:2012tt}.}

{Within such a pionless framework, we} investigated the $PC=++$ states { observed in experiments} with masses about 3.9~GeV, { including the $X(3915)$, the $Z(3930)$, the $\chi_{c0}(3930)$,  the $\chi_{c2}(3930)$, and the $X(3960)$.} We assumed that these structures are due to only two states. Namely, on the one hand, we assumed that the $X(3915)$ found in the $J/\psi\omega$ mode is the same $2^{++}$ state as the  $\chi_{c2}(3930)$~\cite{Zhou:2015uva}, { which is parameterized as a Breit-Wigner resonance in this work, and on the other hand, the isoscalar $D\bar D$-$D_s\bar D_s$-$D^*\bar D^*$-$D_s^*\bar D^*_s$ coupled channels with $0^{++}$ quantum numbers resulted in a molecular state near the $D_s^+D_s^-$ threshold, which accounts for the structures of the $\chi_{c0}(3930)$, the $X(3960)$, and the $0^{++}$ component in the $Z(3930)$.}
We found that both the Belle~\cite{Belle:2005rte} and BaBar~\cite{BaBar:2010jfn} data from the $\gamma\gamma\to D\bar D$ fusion reaction, and the LHCb $H\bar H$ distributions from the  $B^+\to K^+D^+ D^-$~\cite{LHCb:2020pxc} and $B^+\to K^+D_s^+ D_s^-$~\cite{LHCb:2022vsv} decays can be well described simultaneously. 
The results suggest that the identification of the $0^{++}$ component in the LHCb analysis of the  $B^+\to K^+D^+ D^-$~\cite{LHCb:2020pxc} to the $X(3915)$, and correspondingly the assignment of the $X(3915)$ quantum numbers to be $0^{++}$ in the current version of the RPP~\cite{ParticleDataGroup:2022pth} are premature.
Within our analysis, the $0^{++}$ component comes from the $D_s^+D_s^-$ molecular state, which leads to the near-threshold enhancement of the $D_s^+D_s^-$ invariant mass distribution in the $B^+\to K^+D_s^+ D_s^-$ decay.
To test our proposal, we predict that there should be a peak and a dip around the $D_s^+D_s^-$ threshold in the $D\bar D$ invariant mass distributions of the $e^+e^-\to \phi D\bar D$ and $e^+e^-\to \omega D\bar D$ reactions, respectively.
The prediction can be checked at the upcoming upgrade of the Beijing Electron-Positron Collider II~\cite{BESIII:2022mxl} and the possible Super Tau-Charm Facilities~\cite{Guo:2022kdi,Charm-TauFactory:2013cnj}. {The lepto- and photo-production of these states could also contribute valuable information \cite{Yang:2021jof,Albaladejo:2020tzt,Winney:2022tky}.}
At last, we would like to emphasize that the $PC=++$ states in the region from about 3.9 GeV to 4 GeV should be understood together, and HQSS plays a valuable role~\cite{Hidalgo-Duque:2012rqv,Ji:2022uie}.

\section*{Acknowledgements}
We are grateful to the fruitful discussions with Albert Feijoo, Mikhail Mikhasenko, Eulogio Oset, Jia-Jun Wu and Zhen-Hua Zhang. This research has been supported  by the Spanish Ministerio de Ciencia e Innovaci\'on (MICINN)
and the European Regional Development Fund (ERDF) under Contract PID2020-112777GB-I00; 
by the EU STRONG-2020 Project under the Program H2020-INFRAIA-2018-1 with 
Grant Agreement No. 824093; by  Generalitat Valenciana under Contract PROMETEO/2020/023; by the Chinese Academy of Sciences under Grant No.~XDB34030000; by the National Natural Science Foundation of China (NSFC) under Grants No.~12125507, No. 11835015, No.~12047503, and No.~11961141012; and by the NSFC and the Deutsche Forschungsgemeinschaft (DFG) through the funds provided to the Sino-German Collaborative Research Center TRR110 ``Symmetries and the Emergence of Structure in QCD'' (NSFC Grant No.~12070131001, DFG Project-ID~196253076). M.~A. is supported by Generalitat Valenciana under Grant No. CIDEGENT/2020/002.

\vspace{0.3cm}
\appendix

\section{Fit parameters and line shapes without averaging over energy bins}
\label{app:para}

We show in Fig.~\ref{fig:fit-lineshape-noint} the comparison of the best fit results  with  data without averaging over each energy bin.

The central values and errors of the free parameters from the best fit are
\begin{align}
     \mathcal C_{1X}  &=(-0.33\pm0.02)\ \rm fm^2,\notag\\
     \mathcal C_{0a}  &=(-1.36\pm0.06)\ \rm fm^2,\notag\\
     \mathcal C_{1a}  &=(-0.33\pm0.02)\ \rm fm^2,\notag\\
     P_1^+  &=-16\pm 4,\notag\\
     P_1^0/P_1^+  &=-20\pm6,\notag\\
     P_2/P_1^+  &=-3.1\pm2.8,\notag\\
     P_3/P_1^+  &=-19\pm8,\notag\\
     P_4/P_1^+  &=-43\pm8,\notag\\
     P_{\gamma}  &=(51\pm4)\times10^{3},\notag\\
     H_{\gamma0}/2\pi  &=(7.5\pm2.6)\times10^{3},\notag\\ 
     H_{\gamma2}/H_{\gamma0}   &=0.8\pm 0.8,\notag\\ 
     H_{D}  &=(1.1\pm0.1)\times10^3,\notag\\
     H_{S2}  &=15\pm6,\notag\\
     H_{\rm bg}/2\pi  &=(1.13\pm0.11)\times10^{6},\notag\\
     m_2  &=(3922\pm2)\, \rm MeV,\notag\\
     \Gamma_2  &=(16\pm3)\ \rm MeV,\notag\\
     m_0  &=(3815\pm10)\ \rm MeV,\notag\\
     \Gamma_0  &=(90\pm11)\ \rm MeV,\notag\\
     \alpha  &=2.26\pm0.20,\notag\\
     f_{\rm BaBar} &=1.32\pm0.09.\notag
\end{align}
and their statistical correlations are collected in Table~\ref{tab:corr}.

\begin{figure*}[tbh]
    \centering
    \includegraphics[width=\linewidth]{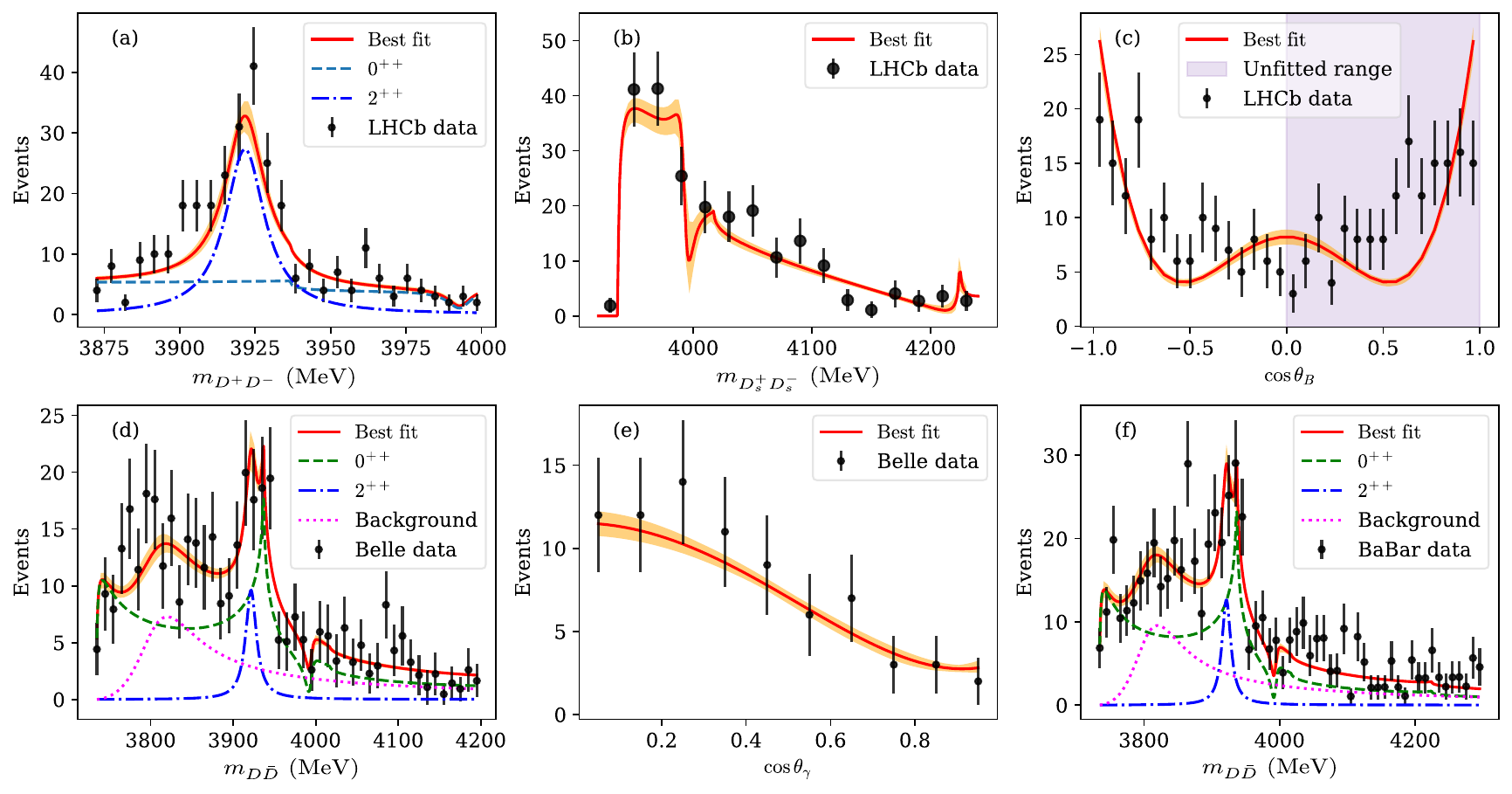}
    \caption{Best fit to the LHCb data for the $B^+\to K^+D^+D^-$~\cite{LHCb:2020pxc}, $B^+\to K^+D_s^+D_s^-$~\cite{LHCb:2022vsv} decays and the Belle~\cite{Belle:2005rte} and BaBar~\cite{BaBar:2010jfn} data for the $\gamma\gamma\to D\bar D$ reaction. The orange bands stand for the statistical uncertainty of the fitted line shapes, which are inherited from the errors of the different data sets. 
    }
    \label{fig:fit-lineshape-noint}
\end{figure*}

\bibliography{ref.bib}

\begin{sidewaystable*}[h]
\centering
\caption{Correlation matrix of the fitted parameters from the best fit.}\label{tab:corr}\footnotesize

\renewcommand\arraystretch{1.2}
\begin{tabular}{rrrrrrrrrrrrrrrrrrrrr}
& \makecell[c]{$\mathcal C_{1X}$}&\makecell[c]{$\mathcal C_{0a}$}&\makecell[c]{$\mathcal C_{1a}$}&\makecell[c]{$ P_1^+$}&\makecell[c]{$ P_1^0/P_1^+$}&\makecell[c]{$P_2/P_1^+$}&\makecell[c]{$P_3/P_1^+$}&\makecell[c]{$ P_4/P_1^+$}&\makecell[c]{$P_{\gamma}$}&\makecell[c]{$H_{\gamma0}$}&\makecell[c]{$H_{\gamma2}/H_{\gamma0}$}&\makecell[c]{$H_{D}$}&\makecell[c]{$H_{S2}$}&\makecell[c]{$H_{\rm bg}$}&$m_2$&\makecell[c]{$ \Gamma_2$}&\makecell[c]{$m_0$}&\makecell[c]{$\Gamma_0$}&\makecell[c]{$\alpha$}&\makecell[c]{$f_{\rm BaBar}$}\\
 \makecell[c]{$\mathcal C_{1X}$}&\cellcolor{poscolor!100.0}$1.0$&\cellcolor{negcolor!7.770205918698084}$-0.08$&\cellcolor{poscolor!97.21305991477647}$0.97$&\cellcolor{negcolor!47.40256273594235}$-0.47$&\cellcolor{poscolor!56.24160822481286}$0.56$&\cellcolor{poscolor!43.54373808799977}$0.44$&\cellcolor{negcolor!61.566694158259374}$-0.62$&\cellcolor{negcolor!65.67960577050606}$-0.66$&\cellcolor{poscolor!6.602385803316286}$0.07$&\cellcolor{negcolor!5.451763121439435}$-0.05$&\cellcolor{poscolor!16.1160552589404}$0.16$&\cellcolor{poscolor!2.2549425159117926}$0.02$&\cellcolor{negcolor!42.22596798749777}$-0.42$&\cellcolor{poscolor!8.628900720695858}$0.09$&\cellcolor{poscolor!4.779371687666636}$0.05$&\cellcolor{negcolor!2.480991928636818}$-0.02$&\cellcolor{negcolor!3.8867168151028966}$-0.04$&\cellcolor{negcolor!29.052509059181986}$-0.29$&\cellcolor{negcolor!15.484817244201382}$-0.15$&\cellcolor{negcolor!6.977569504532344}$-0.07$\\
\makecell[c]{$\mathcal C_{0a}$}&\cellcolor{negcolor!7.770205918698084}$-0.08$&\cellcolor{poscolor!100.0}$1.0$&\cellcolor{negcolor!7.85676068326472}$-0.08$&\cellcolor{poscolor!11.74820267287258}$0.12$&\cellcolor{negcolor!5.3242436248727865}$-0.05$&\cellcolor{negcolor!1.4601536352339282}$-0.01$&\cellcolor{negcolor!0.9680229604810137}$-0.01$&\cellcolor{negcolor!15.83869666431238}$-0.16$&\cellcolor{negcolor!53.778691038253754}$-0.54$&\cellcolor{negcolor!4.129349188091436}$-0.04$&\cellcolor{poscolor!2.4752219634556187}$0.02$&\cellcolor{negcolor!9.406432267813507}$-0.09$&\cellcolor{poscolor!10.777581732785203}$0.11$&\cellcolor{poscolor!27.323370668159253}$0.27$&\cellcolor{negcolor!10.514228576139935}$-0.11$&\cellcolor{poscolor!21.72670781418811}$0.22$&\cellcolor{negcolor!13.857073565061556}$-0.14$&\cellcolor{negcolor!1.4500716410139396}$-0.01$&\cellcolor{poscolor!2.3136673456109675}$0.02$&\cellcolor{negcolor!20.944008765258907}$-0.21$\\
\makecell[c]{$\mathcal C_{1a}$}&\cellcolor{poscolor!97.21305991477647}$0.97$&\cellcolor{negcolor!7.85676068326472}$-0.08$&\cellcolor{poscolor!100.0}$1.0$&\cellcolor{negcolor!54.2980729984358}$-0.54$&\cellcolor{poscolor!62.97245278128095}$0.63$&\cellcolor{poscolor!51.10633401762945}$0.51$&\cellcolor{negcolor!66.55857179911663}$-0.67$&\cellcolor{negcolor!70.01216095100504}$-0.7$&\cellcolor{poscolor!6.137290809273875}$0.06$&\cellcolor{negcolor!5.159867863538392}$-0.05$&\cellcolor{poscolor!16.16020665421491}$0.16$&\cellcolor{poscolor!2.9539657667388384}$0.03$&\cellcolor{negcolor!49.97485668249852}$-0.5$&\cellcolor{poscolor!8.747403540981608}$0.09$&\cellcolor{poscolor!6.626849470872494}$0.07$&\cellcolor{negcolor!3.6735424455805012}$-0.04$&\cellcolor{negcolor!3.66020435968758}$-0.04$&\cellcolor{negcolor!29.58021413234079}$-0.3$&\cellcolor{negcolor!15.973991347565883}$-0.16$&\cellcolor{negcolor!6.843538262584802}$-0.07$\\
\makecell[c]{$ P_1^+$}&\cellcolor{negcolor!47.40256273594235}$-0.47$&\cellcolor{poscolor!11.74820267287258}$0.12$&\cellcolor{negcolor!54.2980729984358}$-0.54$&\cellcolor{poscolor!100.0}$1.0$&\cellcolor{negcolor!96.72605816350317}$-0.97$&\cellcolor{negcolor!94.82259651045662}$-0.95$&\cellcolor{poscolor!89.02968547590693}$0.89$&\cellcolor{poscolor!71.21001201429586}$0.71$&\cellcolor{negcolor!5.796072918998641}$-0.06$&\cellcolor{poscolor!1.934941390610637}$0.02$&\cellcolor{negcolor!8.902014492247638}$-0.09$&\cellcolor{negcolor!0.4683328385914047}$-0.0$&\cellcolor{poscolor!98.04391021199754}$0.98$&\cellcolor{negcolor!8.0826370431687}$-0.08$&\cellcolor{negcolor!15.466012445405774}$-0.15$&\cellcolor{poscolor!17.724963197396864}$0.18$&\cellcolor{poscolor!2.919981096572118}$0.03$&\cellcolor{poscolor!20.52642583624525}$0.21$&\cellcolor{poscolor!19.539646188750147}$0.2$&\cellcolor{poscolor!1.269943796079319}$0.01$\\
\makecell[c]{$ P_1^0/P_1^+$}&\cellcolor{poscolor!56.24160822481286}$0.56$&\cellcolor{negcolor!5.3242436248727865}$-0.05$&\cellcolor{poscolor!62.97245278128095}$0.63$&\cellcolor{negcolor!96.72605816350317}$-0.97$&\cellcolor{poscolor!100.0}$1.0$&\cellcolor{poscolor!98.15273096747677}$0.98$&\cellcolor{negcolor!96.32638599274698}$-0.96$&\cellcolor{negcolor!83.8226227480555}$-0.84$&\cellcolor{negcolor!0.13599040090870446}$-0.0$&\cellcolor{negcolor!3.0337150183703585}$-0.03$&\cellcolor{poscolor!10.290043306216292}$0.1$&\cellcolor{poscolor!4.491618202562404}$0.04$&\cellcolor{negcolor!94.79574172450404}$-0.95$&\cellcolor{poscolor!5.356337561955877}$0.05$&\cellcolor{poscolor!14.665475471601146}$0.15$&\cellcolor{negcolor!10.65679743381068}$-0.11$&\cellcolor{negcolor!1.6496290042455861}$-0.02$&\cellcolor{negcolor!15.31228673259887}$-0.15$&\cellcolor{negcolor!18.408620930264178}$-0.18$&\cellcolor{poscolor!0.756209426199055}$0.01$\\
\makecell[c]{$P_2/P_1^+$}&\cellcolor{poscolor!43.54373808799977}$0.44$&\cellcolor{negcolor!1.4601536352339282}$-0.01$&\cellcolor{poscolor!51.10633401762945}$0.51$&\cellcolor{negcolor!94.82259651045662}$-0.95$&\cellcolor{poscolor!98.15273096747677}$0.98$&\cellcolor{poscolor!100.0}$1.0$&\cellcolor{negcolor!94.10668943141842}$-0.94$&\cellcolor{negcolor!81.45787393468147}$-0.81$&\cellcolor{negcolor!2.668123979778713}$-0.03$&\cellcolor{negcolor!2.1419675072925326}$-0.02$&\cellcolor{poscolor!8.317740147018707}$0.08$&\cellcolor{poscolor!5.668499148354755}$0.06$&\cellcolor{negcolor!94.59466177847098}$-0.95$&\cellcolor{poscolor!3.2027730102512817}$0.03$&\cellcolor{poscolor!15.607396232889098}$0.16$&\cellcolor{negcolor!10.489612378355137}$-0.1$&\cellcolor{negcolor!0.5675132158411187}$-0.01$&\cellcolor{negcolor!9.714631518578756}$-0.1$&\cellcolor{negcolor!16.63214697648614}$-0.17$&\cellcolor{poscolor!1.6991179134531496}$0.02$\\
\makecell[c]{$P_3/P_1^+$}&\cellcolor{negcolor!61.566694158259374}$-0.62$&\cellcolor{negcolor!0.9680229604810137}$-0.01$&\cellcolor{negcolor!66.55857179911663}$-0.67$&\cellcolor{poscolor!89.02968547590693}$0.89$&\cellcolor{negcolor!96.32638599274698}$-0.96$&\cellcolor{negcolor!94.10668943141842}$-0.94$&\cellcolor{poscolor!100.0}$1.0$&\cellcolor{poscolor!89.16603508648993}$0.89$&\cellcolor{poscolor!3.297457917145674}$0.03$&\cellcolor{poscolor!2.6376603416719426}$0.03$&\cellcolor{negcolor!10.80365712907839}$-0.11$&\cellcolor{negcolor!6.9040442012555685}$-0.07$&\cellcolor{poscolor!86.50954008005057}$0.87$&\cellcolor{negcolor!2.5313679773497313}$-0.03$&\cellcolor{negcolor!14.51548643202542}$-0.15$&\cellcolor{poscolor!8.722376566772134}$0.09$&\cellcolor{poscolor!0.6028400059729826}$0.01$&\cellcolor{poscolor!11.2274036264692}$0.11$&\cellcolor{poscolor!18.024472315691234}$0.18$&\cellcolor{negcolor!0.13203282989417148}$-0.0$\\
\makecell[c]{$ P_4/P_1^+$}&\cellcolor{negcolor!65.67960577050606}$-0.66$&\cellcolor{negcolor!15.83869666431238}$-0.16$&\cellcolor{negcolor!70.01216095100504}$-0.7$&\cellcolor{poscolor!71.21001201429586}$0.71$&\cellcolor{negcolor!83.8226227480555}$-0.84$&\cellcolor{negcolor!81.45787393468147}$-0.81$&\cellcolor{poscolor!89.16603508648993}$0.89$&\cellcolor{poscolor!100.0}$1.0$&\cellcolor{poscolor!9.389297577270161}$0.09$&\cellcolor{poscolor!3.5858465991404116}$0.04$&\cellcolor{negcolor!12.171186999722948}$-0.12$&\cellcolor{negcolor!7.640879808102594}$-0.08$&\cellcolor{poscolor!68.48115289002871}$0.68$&\cellcolor{negcolor!4.727243474184909}$-0.05$&\cellcolor{negcolor!11.088367620967748}$-0.11$&\cellcolor{poscolor!5.545985548631034}$0.06$&\cellcolor{poscolor!2.050007712409373}$0.02$&\cellcolor{poscolor!12.296975895763328}$0.12$&\cellcolor{poscolor!20.998037114413215}$0.21$&\cellcolor{poscolor!4.606912338146863}$0.05$\\
\makecell[c]{$P_{\gamma}$}&\cellcolor{poscolor!6.602385803316286}$0.07$&\cellcolor{negcolor!53.778691038253754}$-0.54$&\cellcolor{poscolor!6.137290809273875}$0.06$&\cellcolor{negcolor!5.796072918998641}$-0.06$&\cellcolor{negcolor!0.13599040090870446}$-0.0$&\cellcolor{negcolor!2.668123979778713}$-0.03$&\cellcolor{poscolor!3.297457917145674}$0.03$&\cellcolor{poscolor!9.389297577270161}$0.09$&\cellcolor{poscolor!100.0}$1.0$&\cellcolor{poscolor!14.528449972551599}$0.15$&\cellcolor{negcolor!23.785870057777952}$-0.24$&\cellcolor{poscolor!13.833137366179137}$0.14$&\cellcolor{negcolor!7.06855239763102}$-0.07$&\cellcolor{negcolor!61.95745770410004}$-0.62$&\cellcolor{poscolor!9.976986067195307}$0.1$&\cellcolor{negcolor!28.013248811464564}$-0.28$&\cellcolor{poscolor!29.126138735197284}$0.29$&\cellcolor{negcolor!9.943508618488515}$-0.1$&\cellcolor{negcolor!4.458559516485007}$-0.04$&\cellcolor{negcolor!14.005176867011038}$-0.14$\\
\makecell[c]{$H_{\gamma0}$}&\cellcolor{negcolor!5.451763121439435}$-0.05$&\cellcolor{negcolor!4.129349188091436}$-0.04$&\cellcolor{negcolor!5.159867863538392}$-0.05$&\cellcolor{poscolor!1.934941390610637}$0.02$&\cellcolor{negcolor!3.0337150183703585}$-0.03$&\cellcolor{negcolor!2.1419675072925326}$-0.02$&\cellcolor{poscolor!2.6376603416719426}$0.03$&\cellcolor{poscolor!3.5858465991404116}$0.04$&\cellcolor{poscolor!14.528449972551599}$0.15$&\cellcolor{poscolor!100.0}$1.0$&\cellcolor{negcolor!91.4954212870958}$-0.91$&\cellcolor{poscolor!5.935103534543764}$0.06$&\cellcolor{poscolor!1.1978887050386129}$0.01$&\cellcolor{negcolor!3.571997468594385}$-0.04$&\cellcolor{negcolor!1.4231499826441627}$-0.01$&\cellcolor{negcolor!8.947764574244466}$-0.09$&\cellcolor{poscolor!2.0965844242647496}$0.02$&\cellcolor{negcolor!5.6632679037397695}$-0.06$&\cellcolor{negcolor!0.38647496495381467}$-0.0$&\cellcolor{negcolor!5.884640624911158}$-0.06$\\
\makecell[c]{$H_{\gamma2}/H_{\gamma0}$}&\cellcolor{poscolor!16.1160552589404}$0.16$&\cellcolor{poscolor!2.4752219634556187}$0.02$&\cellcolor{poscolor!16.16020665421491}$0.16$&\cellcolor{negcolor!8.902014492247638}$-0.09$&\cellcolor{poscolor!10.290043306216292}$0.1$&\cellcolor{poscolor!8.317740147018707}$0.08$&\cellcolor{negcolor!10.80365712907839}$-0.11$&\cellcolor{negcolor!12.171186999722948}$-0.12$&\cellcolor{negcolor!23.785870057777952}$-0.24$&\cellcolor{negcolor!91.4954212870958}$-0.91$&\cellcolor{poscolor!100.0}$1.0$&\cellcolor{poscolor!1.1348749213831817}$0.01$&\cellcolor{negcolor!8.541546541886952}$-0.09$&\cellcolor{poscolor!15.991954464723337}$0.16$&\cellcolor{poscolor!4.231756216171451}$0.04$&\cellcolor{negcolor!0.9270678083611367}$-0.01$&\cellcolor{negcolor!12.990271252065305}$-0.13$&\cellcolor{negcolor!5.592390819894741}$-0.06$&\cellcolor{negcolor!3.1473893246169387}$-0.03$&\cellcolor{poscolor!1.510456903476197}$0.02$\\
\makecell[c]{$H_{D}$}&\cellcolor{poscolor!2.2549425159117926}$0.02$&\cellcolor{negcolor!9.406432267813507}$-0.09$&\cellcolor{poscolor!2.9539657667388384}$0.03$&\cellcolor{negcolor!0.4683328385914047}$-0.0$&\cellcolor{poscolor!4.491618202562404}$0.04$&\cellcolor{poscolor!5.668499148354755}$0.06$&\cellcolor{negcolor!6.9040442012555685}$-0.07$&\cellcolor{negcolor!7.640879808102594}$-0.08$&\cellcolor{poscolor!13.833137366179137}$0.14$&\cellcolor{poscolor!5.935103534543764}$0.06$&\cellcolor{poscolor!1.1348749213831817}$0.01$&\cellcolor{poscolor!100.0}$1.0$&\cellcolor{negcolor!0.9925898041803662}$-0.01$&\cellcolor{poscolor!5.330490079846728}$0.05$&\cellcolor{poscolor!32.13539995349892}$0.32$&\cellcolor{negcolor!70.36599747602669}$-0.7$&\cellcolor{poscolor!0.6100727272492386}$0.01$&\cellcolor{negcolor!13.246803471729434}$-0.13$&\cellcolor{poscolor!1.5367422336513175}$0.02$&\cellcolor{negcolor!3.6476348762174022}$-0.04$\\
\makecell[c]{$H_{S2}$}&\cellcolor{negcolor!42.22596798749777}$-0.42$&\cellcolor{poscolor!10.777581732785203}$0.11$&\cellcolor{negcolor!49.97485668249852}$-0.5$&\cellcolor{poscolor!98.04391021199754}$0.98$&\cellcolor{negcolor!94.79574172450404}$-0.95$&\cellcolor{negcolor!94.59466177847098}$-0.95$&\cellcolor{poscolor!86.50954008005057}$0.87$&\cellcolor{poscolor!68.48115289002871}$0.68$&\cellcolor{negcolor!7.06855239763102}$-0.07$&\cellcolor{poscolor!1.1978887050386129}$0.01$&\cellcolor{negcolor!8.541546541886952}$-0.09$&\cellcolor{negcolor!0.9925898041803662}$-0.01$&\cellcolor{poscolor!100.0}$1.0$&\cellcolor{negcolor!9.678902929819069}$-0.1$&\cellcolor{negcolor!16.62172242306573}$-0.17$&\cellcolor{poscolor!18.55404161789523}$0.19$&\cellcolor{poscolor!3.8938266445911855}$0.04$&\cellcolor{poscolor!22.314917244610218}$0.22$&\cellcolor{poscolor!18.61321180361196}$0.19$&\cellcolor{poscolor!4.701063985545052}$0.05$\\
\makecell[c]{$H_{\rm bg}$}&\cellcolor{poscolor!8.628900720695858}$0.09$&\cellcolor{poscolor!27.323370668159253}$0.27$&\cellcolor{poscolor!8.747403540981608}$0.09$&\cellcolor{negcolor!8.0826370431687}$-0.08$&\cellcolor{poscolor!5.356337561955877}$0.05$&\cellcolor{poscolor!3.2027730102512817}$0.03$&\cellcolor{negcolor!2.5313679773497313}$-0.03$&\cellcolor{negcolor!4.727243474184909}$-0.05$&\cellcolor{negcolor!61.95745770410004}$-0.62$&\cellcolor{negcolor!3.571997468594385}$-0.04$&\cellcolor{poscolor!15.991954464723337}$0.16$&\cellcolor{poscolor!5.330490079846728}$0.05$&\cellcolor{negcolor!9.678902929819069}$-0.1$&\cellcolor{poscolor!100.0}$1.0$&\cellcolor{poscolor!9.958799776790398}$0.1$&\cellcolor{negcolor!13.02439530142866}$-0.13$&\cellcolor{negcolor!35.71097487984273}$-0.36$&\cellcolor{negcolor!34.56871646082469}$-0.35$&\cellcolor{negcolor!4.037422888295362}$-0.04$&\cellcolor{negcolor!21.527983989741344}$-0.22$\\
\makecell[c]{$m_2$}&\cellcolor{poscolor!4.779371687666636}$0.05$&\cellcolor{negcolor!10.514228576139935}$-0.11$&\cellcolor{poscolor!6.626849470872494}$0.07$&\cellcolor{negcolor!15.466012445405774}$-0.15$&\cellcolor{poscolor!14.665475471601146}$0.15$&\cellcolor{poscolor!15.607396232889098}$0.16$&\cellcolor{negcolor!14.51548643202542}$-0.15$&\cellcolor{negcolor!11.088367620967748}$-0.11$&\cellcolor{poscolor!9.976986067195307}$0.1$&\cellcolor{negcolor!1.4231499826441627}$-0.01$&\cellcolor{poscolor!4.231756216171451}$0.04$&\cellcolor{poscolor!32.13539995349892}$0.32$&\cellcolor{negcolor!16.62172242306573}$-0.17$&\cellcolor{poscolor!9.958799776790398}$0.1$&\cellcolor{poscolor!100.0}$1.0$&\cellcolor{negcolor!39.919052391798004}$-0.4$&\cellcolor{negcolor!3.5993856882186517}$-0.04$&\cellcolor{negcolor!16.293128633493957}$-0.16$&\cellcolor{poscolor!5.922521424517595}$0.06$&\cellcolor{negcolor!4.4139108074383255}$-0.04$\\
\makecell[c]{$ \Gamma_2$}&\cellcolor{negcolor!2.480991928636818}$-0.02$&\cellcolor{poscolor!21.72670781418811}$0.22$&\cellcolor{negcolor!3.6735424455805012}$-0.04$&\cellcolor{poscolor!17.724963197396864}$0.18$&\cellcolor{negcolor!10.65679743381068}$-0.11$&\cellcolor{negcolor!10.489612378355137}$-0.1$&\cellcolor{poscolor!8.722376566772134}$0.09$&\cellcolor{poscolor!5.545985548631034}$0.06$&\cellcolor{negcolor!28.013248811464564}$-0.28$&\cellcolor{negcolor!8.947764574244466}$-0.09$&\cellcolor{negcolor!0.9270678083611367}$-0.01$&\cellcolor{negcolor!70.36599747602669}$-0.7$&\cellcolor{poscolor!18.55404161789523}$0.19$&\cellcolor{negcolor!13.02439530142866}$-0.13$&\cellcolor{negcolor!39.919052391798004}$-0.4$&\cellcolor{poscolor!100.0}$1.0$&\cellcolor{poscolor!2.1390858250730838}$0.02$&\cellcolor{poscolor!31.258810817701903}$0.31$&\cellcolor{poscolor!10.967709506198327}$0.11$&\cellcolor{poscolor!10.996263050714614}$0.11$\\
\makecell[c]{$m_0$}&\cellcolor{negcolor!3.8867168151028966}$-0.04$&\cellcolor{negcolor!13.857073565061556}$-0.14$&\cellcolor{negcolor!3.66020435968758}$-0.04$&\cellcolor{poscolor!2.919981096572118}$0.03$&\cellcolor{negcolor!1.6496290042455861}$-0.02$&\cellcolor{negcolor!0.5675132158411187}$-0.01$&\cellcolor{poscolor!0.6028400059729826}$0.01$&\cellcolor{poscolor!2.050007712409373}$0.02$&\cellcolor{poscolor!29.126138735197284}$0.29$&\cellcolor{poscolor!2.0965844242647496}$0.02$&\cellcolor{negcolor!12.990271252065305}$-0.13$&\cellcolor{poscolor!0.6100727272492386}$0.01$&\cellcolor{poscolor!3.8938266445911855}$0.04$&\cellcolor{negcolor!35.71097487984273}$-0.36$&\cellcolor{negcolor!3.5993856882186517}$-0.04$&\cellcolor{poscolor!2.1390858250730838}$0.02$&\cellcolor{poscolor!100.0}$1.0$&\cellcolor{poscolor!16.150091032081413}$0.16$&\cellcolor{poscolor!2.329422274394085}$0.02$&\cellcolor{poscolor!9.897803422974146}$0.1$\\
\makecell[c]{$\Gamma_0$}&\cellcolor{negcolor!29.052509059181986}$-0.29$&\cellcolor{negcolor!1.4500716410139396}$-0.01$&\cellcolor{negcolor!29.58021413234079}$-0.3$&\cellcolor{poscolor!20.52642583624525}$0.21$&\cellcolor{negcolor!15.31228673259887}$-0.15$&\cellcolor{negcolor!9.714631518578756}$-0.1$&\cellcolor{poscolor!11.2274036264692}$0.11$&\cellcolor{poscolor!12.296975895763328}$0.12$&\cellcolor{negcolor!9.943508618488515}$-0.1$&\cellcolor{negcolor!5.6632679037397695}$-0.06$&\cellcolor{negcolor!5.592390819894741}$-0.06$&\cellcolor{negcolor!13.246803471729434}$-0.13$&\cellcolor{poscolor!22.314917244610218}$0.22$&\cellcolor{negcolor!34.56871646082469}$-0.35$&\cellcolor{negcolor!16.293128633493957}$-0.16$&\cellcolor{poscolor!31.258810817701903}$0.31$&\cellcolor{poscolor!16.150091032081413}$0.16$&\cellcolor{poscolor!100.0}$1.0$&\cellcolor{poscolor!10.505828245575655}$0.11$&\cellcolor{negcolor!9.77974952335763}$-0.1$\\
\makecell[c]{$\alpha$}&\cellcolor{negcolor!15.484817244201382}$-0.15$&\cellcolor{poscolor!2.3136673456109675}$0.02$&\cellcolor{negcolor!15.973991347565883}$-0.16$&\cellcolor{poscolor!19.539646188750147}$0.2$&\cellcolor{negcolor!18.408620930264178}$-0.18$&\cellcolor{negcolor!16.63214697648614}$-0.17$&\cellcolor{poscolor!18.024472315691234}$0.18$&\cellcolor{poscolor!20.998037114413215}$0.21$&\cellcolor{negcolor!4.458559516485007}$-0.04$&\cellcolor{negcolor!0.38647496495381467}$-0.0$&\cellcolor{negcolor!3.1473893246169387}$-0.03$&\cellcolor{poscolor!1.5367422336513175}$0.02$&\cellcolor{poscolor!18.61321180361196}$0.19$&\cellcolor{negcolor!4.037422888295362}$-0.04$&\cellcolor{poscolor!5.922521424517595}$0.06$&\cellcolor{poscolor!10.967709506198327}$0.11$&\cellcolor{poscolor!2.329422274394085}$0.02$&\cellcolor{poscolor!10.505828245575655}$0.11$&\cellcolor{poscolor!100.0}$1.0$&\cellcolor{poscolor!2.659617778887316}$0.03$\\
\makecell[c]{$f_{\rm BaBar}$}&\cellcolor{negcolor!6.977569504532344}$-0.07$&\cellcolor{negcolor!20.944008765258907}$-0.21$&\cellcolor{negcolor!6.843538262584802}$-0.07$&\cellcolor{poscolor!1.269943796079319}$0.01$&\cellcolor{poscolor!0.756209426199055}$0.01$&\cellcolor{poscolor!1.6991179134531496}$0.02$&\cellcolor{negcolor!0.13203282989417148}$-0.0$&\cellcolor{poscolor!4.606912338146863}$0.05$&\cellcolor{negcolor!14.005176867011038}$-0.14$&\cellcolor{negcolor!5.884640624911158}$-0.06$&\cellcolor{poscolor!1.510456903476197}$0.02$&\cellcolor{negcolor!3.6476348762174022}$-0.04$&\cellcolor{poscolor!4.701063985545052}$0.05$&\cellcolor{negcolor!21.527983989741344}$-0.22$&\cellcolor{negcolor!4.4139108074383255}$-0.04$&\cellcolor{poscolor!10.996263050714614}$0.11$&\cellcolor{poscolor!9.897803422974146}$0.1$&\cellcolor{negcolor!9.77974952335763}$-0.1$&\cellcolor{poscolor!2.659617778887316}$0.03$&\cellcolor{poscolor!100.0}$1.0$

\end{tabular}
\end{sidewaystable*}
 
\end{document}